\documentclass[aps,prd,twocolumn,floatfix,noshowpacs,tightenlines,noshowkeys,superscriptaddress,amsmath,amssymb,
nofootinbib]{revtex4-2}

\usepackage[normalem]{ulem}
\usepackage[dvipsnames]{xcolor}
\usepackage[colorlinks,urlcolor=NavyBlue,citecolor=NavyBlue,linkcolor=NavyBlue,pdfusetitle]{hyperref}
\usepackage[all]{hypcap}
\usepackage{enumerate}

\usepackage{graphicx}
\usepackage{comment}
\usepackage{xcolor}
\usepackage{mathrsfs}

\newcommand{\mmin}{m_{\rm min}}
\newcommand{\mmax}{m_{\rm max}}

\newcommand{\nocontentsline}[3]{}
\newcommand{\tocless}[2]{\bgroup\let\addcontentsline=\nocontentsline#1{#2}\egroup}

\newcommand{\hu}{\,{\rm km \, Mpc^{-1}\,s^{-1}}} 

\begin{document}

\title{Cosmology in the dark: On the importance of source population models for gravitational-wave cosmology}

\author{S.~Mastrogiovanni}
\affiliation{Universit\'e de Paris, CNRS, AstroParticule et Cosmologie (APC), F-75013 Paris, France}
\author{K.~Leyde}
\affiliation{Universit\'e de Paris, CNRS, AstroParticule et Cosmologie (APC), F-75013 Paris, France}
\author{C.~Karathanasis}
\affiliation{Institut de F\'isica d'Altes Energies (IFAE), Barcelona Institute of Science and Technology, Barcelona, Spain}
\author{E.~Chassande-Mottin}
\affiliation{Universit\'e de Paris, CNRS, AstroParticule et Cosmologie (APC), F-75013 Paris, France}
\author{D.~A.~Steer}
\affiliation{Universit\'e de Paris, CNRS, AstroParticule et Cosmologie (APC), F-75013 Paris, France}
\author{J.~Gair}
\affiliation{Max Planck Institute for Gravitational Physics (Albert Einstein Institute),
Am M\"uhlenberg 1, Potsdam 14476, Germany}
\author{A.~Ghosh}
\affiliation{Ghent University, Proeftuinstraat 86, 9000 Gent, Belgium}
\author{R.~Gray}
\affiliation{SUPA, University of Glasgow, Glasgow G12 8QQ, United Kingdom}
\author{S.~Mukherjee}
\affiliation{Gravitation Astroparticle Physics Amsterdam (GRAPPA), Anton Pannekoek Institute for Astronomy and Institute for Physics, 
University of Amsterdam, Science Park 904, 1090 GL Amsterdam, The Netherlands}
\affiliation{ Institute Lorentz, Leiden University, PO Box 9506, Leiden 2300 RA, The Netherlands}
\affiliation{Delta Institute for Theoretical Physics, Science Park 904, 1090 GL Amsterdam, The Netherlands}
\author{S.~Rinaldi}
\affiliation{Dipartimento di Fisica "E. Fermi", Università di Pisa, I-56127 Pisa, Italy}
\affiliation{INFN, Sezione di Pisa, I-56127 Pisa, Italy}
\date{\today}

\begin{abstract}
Knowledge of the shape of the mass spectrum of compact objects can be used to help break the degeneracy between the mass and redshift of the gravitational wave (GW) sources, and thus can be used to infer cosmological parameters in the absence of redshift measurements obtained from electromagnetic observations. In this paper we study extensively different aspects of this approach, including its computational limits and achievable accuracy.  
Focusing on ground-based detectors with current and future sensitivities, we first perform the analysis of an extensive set of simulated data using a hierarchical Bayesian scheme that jointly fits the source population and cosmological parameters. We consider a population model (power-law plus Gaussian) which exhibits characteristic scales (extremes of the mass spectrum, presence of an accumulation point modelled by a Gaussian peak) that allow an indirect estimate of the source redshift. Our analysis of this catalogue highlights and quantifies the tight interplay between source population and cosmological parameters, as well as the influence of initial assumptions (whether formulated on source or cosmological parameters).  We then validate our results by an ``end-to-end'' analysis using  simulated GW $h(t)$ data and posterior samples generated from Bayesian samplers used for GW parameter estimation, thus mirroring the analysis chain used for observational data for the first time in literature. Our results then lead us to re-examine the estimation of $H_0$ obtained with GWTC-1 in \cite{Abbott:2019yzh}, and we show explicitly how population assumptions impact the final $H_0$ result.
Together, our results underline the importance of inferring source population and cosmological parameters simultaneously (and not separately as is often assumed).  The only exception, as we discuss, is {\it{if}} an electromagnetic counterpart were to be observed for all the BBH events: then the population assumptions have less impact on the estimation of cosmological parameters.
\end{abstract}

\maketitle

{\footnotesize \setcounter{tocdepth}{0} \tableofcontents}

\section{Introduction}

Gravitational waves (GWs) \cite{Abbott:2016blz,LIGOScientific:2018mvr} from compact binary mergers are often referred to as ``\textit{standard sirens}'', in analogy with the term ``standard candles'' coined for SNIa, thus underlining their role for cosmology. From the GW signal it is possible to directly estimate the source luminosity distance $d_L$  \cite{Sathyaprakash:2009xs,Holz:2005df}. When combined with the redshift of the host galaxy, this estimate can be used to measure cosmological parameters and thus probe the expansion history of the universe.

Probing the expansion history of the universe is crucial to resolve open issues in the standard cosmological model, such as the nature of dark energy and the tension in the values of the Hubble constant $H_0$ i.e. the expansion rate of the Universe today, obtained from observations at early and late cosmological epochs  \cite{Riess:2016jrr,Aghanim:2018eyx,Freedman:2017yms,Riess:2019cxk}.

GWs detected by the LIGO and Virgo experiments \cite{TheLIGOScientific:2014jea,TheVirgo:2014hva} have been used to infer $H_0$ using various approaches and data sets. A first approach \cite{Schutz:1986,Holz:2005df} is to obtain the source redshift by locating the host galaxy thanks to an electromagnetic counterpart to the GW signal. This approach has so far been applied in two cases. The measurement $H_0=70^{+19}_{-8} \hu$ in \cite{Abbott:2017xzu,Abbott:2018wiz} was obtained after the observation of the \textit{kilonova} optical transient that  allowed the galaxy hosting the binary neutron-star (BNS) GW170817 to be pinpointed \cite{TheLIGOScientific:2017qsa}.
Similarly, the optical transient \cite{Graham:2020gwr} tentatively associated to the binary black hole event GW190521 \cite{Abbott:2020mjq,Abbott:2020tfl} led to  $H_0=48^{+24}_{-10} \hu$ \cite{Chen:2020gek,Mukherjee:2020kki}. From GW sources with electromagnetic counterparts, it is also possible to test the theory of general relativity (GR) through GW propagation effects \cite{2020PhRvD.102d4009M}. In order to make an accurate measurements of cosmological parameters and test GR, it is also important --- and indeed essential for the GW sources situated at low redshift --- to correct for peculiar velocity of galaxies  \cite{Mukherjee:2019qmm, Nicolaou:2019cip}.

A second approach \cite{Schutz:1986} consists in establishing a statistical association between the source, and those galaxies in a catalog that match the source sky location and luminosity distance as inferred from GW data. This is well suited to binary black hole (BBH) mergers, for which electromagnetic counterparts are not expected. (So far, there is no clear and robust discovery of a counterpart). A proof of principle application of this approach was applied to GW170817, ignoring the counterpart, finding $H_0={77}_{-18}^{+37} \hu$\cite{Fishbach:2018gjp}. This approach was also applied to the BBH signals detected during the first and second observing runs of Advanced LIGO and Virgo \cite{LIGOScientific:2018mvr,Soares-Santos:2019irc,Abbott:2019yzh} leading to a value of $H_0 = 69^{+16}_{-8} \hu$, when combined with the BNS counterpart measurement. An analysis of the asymmetric mass ratio event GW190814 \cite{Abbott:2020khf} detected during the first half of observing run 3 \cite{Abbott:2020niy} resulted in
the estimate $H_0=70^{+17}_{-8} \hu$. A more recent result using also O3a events finds $H_0=70^{+11}_{-7} \hu$ \cite{Finke:2021aom}.

Several recent studies characterize the future prospects for both approaches in the context of the upcoming observing runs (O4 and O5) for Advanced LIGO and Advanced Virgo, and for the 3rd generation detectors such as the Einstein Telescope (ET). They all concur that it will be increasingly difficult to obtain reliable and precise redshift measurements from electromagnetic observations. Indeed, as GW detector sensitivities improve, the average distance of the detected events increases, and the search for electromagnetic counterparts becomes more challenging \cite{Mastrogiovanni:2020ppa, Chen:2020zoq}: sources at greater distances have dimmer counterparts and a larger number of potential host galaxies. Also the lack of completeness of galaxy surveys at high redshifts will prevent the statistical counterpart association for a large fraction of BBHs that will be observed by the future GW detectors \cite{Maggiore:2019uih}. 


These limitations have motivated the development of 
alternative methods to obtain the source redshift $z$, for instance by the cross-correlating  GW sources with galaxies, see \cite{Oguri:2016dgk, Mukherjee:2018ebj,Mukherjee:2019wcg, Mukherjee:2020hyn,Mukherjee:2020mha}. Here, however, we consider a  different method using solely on GW data. It is based on assumptions about the masses of the compact stars in the 
{\it source frame}. 
The basic idea is the following: from the GW signal it is possible to infer \textit{redshifted} {\it detector-frame} masses $M_z$, where $M_z= (1+z)M$. Therefore the source redshift $z$ can be deduced from the measured detector-frame mass and a statistical estimate of the source-frame mass based on a belief about its distribution. This requires solid prior knowledge of the mass distribution that can be inferred from available data.
Mass distribution with typical source-frame mass scales associated with accumulation points (narrow peaks) or sudden extinction (sharp breaks), can be used to infer the redshift of those GW events falling close-by, through a comparison with their observed detector-frame mass. 
This idea has been explored in several works, which analyse how one can constrain mass distributions and cosmology together.   

In \cite{Taylor:2011fs} the authors propose exploiting the narrow binary {\it neutron star} component mass distribution (normal distribution with a few percent scatter) \cite{Kiziltan:2013, Valentim:2011} to constrain $H_0$ within $10\%$ using hundreds of LIGO and Virgo GW events. Assuming $H_0$, $\Omega_{m,0}$ and $\Omega_{k,0}$ are known at the sub-percent accuracy, reference \cite{Taylor:2012db} follows the same idea to constrain the equation of state of dark matter from ET observations.

Regarding {\it black holes}, their mass distribution is expected to be shaped by various processes. The pair-instability supernovae (PISN) process \cite{Bond:1984}, is expected to lead to a depletion in BHs with masses from $\sim50$ to $\sim120 M_\odot$, often referred to as the ``mass gap''. These scales can be used to extract cosmological parameters. In \cite{Farr:2019twy} for instance, the authors simulate a population of BBHs with a PISN feature at $45 M_\odot$ showing that with 5 years of Advanced LIGO and Virgo, it will be possible to estimate the Hubble parameter at $z=0.8$ with $\sim6.1\%$ precision, the dark energy equation of state parameter to $\sim 10\%$ accuracy and the location of the PISN feature with $5\%$ accuracy.
Similarly in \cite{You:2020wju}, the authors simulate a population of BBHs showing that after a year of ET observations, $H_0$ is expected to be measured at the percent level when fixing all the population parameters. Ref.~\cite{You:2020wju} also shows that other population related parameters, such as the rate evolution parameter, could be crucial to infer the $H_0$ (even though this simulation is done without using errors on the measurement of the GW signal parameters).  
In \cite{Ezquiaga:2020tns} the authors discuss using the {\it higher end} of the PISN mass gap (i.e., the observation of intermediate-mass black hole binaries) and in the context of ET, they estimate $H_0$ will be determined at $\lesssim 20\%$ accuracy (with this reducing to $\sim 3\%$ in the most optimistic scenario).

In contrast to previous works, which are mainly focused on providing forecasts for the measurement of cosmological parameters, in this paper we study in depth several technical aspects related to this type of analysis. The study concentrates on the near term and the upcoming LIGO and Virgo observing runs by simulating a BBH population similar to the one inferred from \cite{Abbott:2020gyp}.
In Sec.~\ref{sec:2} we summarize the joint inferential scheme for both cosmological and source population parameters. In Sec.~\ref{sec:catalogue1} we apply this scheme to a simulated BBH population, and identify the most important source population parameters for GW cosmology. We also discuss the convergence of the errors on different parameters as a function of the number of detected events.  In Sec.~\ref{sec:3} we study the interplay between mass-population and cosmological parameters, focusing on some cases of particular relevance. In Sec.~\ref{sec:4}, we discuss the effect on the $H_0$ estimation of choosing a different mass model from that of the simulated population. In section \ref{sec:5} validate our results by an ``end-to-end'' analysis using  simulated GW $h(t)$ data and posterior samples generated from Bayesian samplers used for GW parameter estimation, thus mirroring the analysis chain used for observational data for the first time in literature. This leads us, in section \ref{sec:reanalysis} to re-examine the estimation of $H_0$ obtained with GWTC-1 in \cite{Abbott:2019yzh}, and we show explicitly how population assumptions impact the final $H_0$ result. Finally, in section \ref{sec:EM} we show that {\it{if}} an electromagnetic counterpart were to be observed for all the BBH events, then the population assumptions would not impact the estimation of cosmological parameters. Our conclusions are summarized in section \ref{sec:conc}.

\section{Hierarchical Bayesian analysis \label{sec:2}}

In this section, we introduce our notation and outline the scheme for jointly inferring cosmological and source population parameters.

\subsection{Notation and definition of source population models}

We denote by $\theta$ the set of parameters describing individual black-hole sources in the source frame. For the present analysis, the most important amongst these are the source-frame masses, $m_{1/2,s}$, of the two binary components, and the cosmological redshift $z$ (others include the spins, position of the source on the sky, orientation, eccentricity etc). The distribution of BH sources in the population is described by a set of hyper-parameters denoted by $\Lambda_m$, while the cosmological parameters include the Hubble constant $H_0$ and the present-day fraction of matter density $\Omega_{m,0}$ (for a flat $\Lambda$CDM Universe). We denote them by $\Lambda_c=\{H_0,\Omega_{m,0}\}$. Often in the following we will collect all (cosmological and source population) hyper-parameters together and denote them by $\Lambda = \{\Lambda_m, \Lambda_c\}$. 

The distribution of individual source properties $p_{\rm pop}(\theta|\Lambda_m,H_0,\Omega_{m,0})$ is taken to be of the form 
\begin{eqnarray}
 p_{\rm pop}(\theta|\Lambda_m,\Lambda_c) &=& C \: p(m_{1,s},m_{2,s}|\Lambda_m) \nonumber \\ && \times  \frac{dV_c}{dz}(\Lambda_c)(1+z)^{\gamma-1},
 \label{eq:popinduced}
\end{eqnarray}
where $p(m_{1,s},m_{2,s}|\Lambda_m)$ describes the source-frame mass distribution (see below); $\frac{dV_c}{dz} (\Lambda_c)= \frac{4\pi c(1+z)^2D^2_A}{H(z)}$ is the differential comoving volume, with $D_A$ the angular diameter distance and $H(z)$ the Hubble parameter; the factor of $(1+z)$ in Eq.~\eqref{eq:popinduced} is the standard time dilatation between source and detector frame clocks; finally the power-law index $\gamma$ characterizes the merger rate evolution with redshift \cite{Fishbach:2018edt} (a null value of $\gamma$ corresponds to a constant merger rate in comoving volume). Finally the constant $C$ ensures proper normalization of the probability distribution to unity.

We use two models for the source-frame mass spectrum that were previously implemented in \cite{LIGOScientific:2018jsj,Abbott:2020gyp}. The first is simple power-law model, labelled PL, in which the prior on the first component mass $m_{1,s}$ is a power law with index ($-\alpha$) and lower and upper cutoffs at $\mmin$ and $\mmax$ respectively. The second component mass is distributed according to a power law with index $\beta$ between $\mmin$ and $m_{1,s}$. The corresponding explicit source-frame mass distribution is given in Appendix \ref{app:a}.
This simple model is completely determined by the four parameters ($\mmax, \mmin, \alpha,\beta$).

The second more complex model is labelled PLG. Here the first component mass follows the same PL model as above with the addition of a Gaussian peak with mean $\mu_g$ and variance $\sigma_g^2$. The proportion of events that arise from the Gaussian peak is governed by the parameter $\lambda_g$ (when $\lambda_g=0$, the model PLG reduces to PL). The second mass component is drawn as in the previous model. In addition, this model also includes a tapering factor $\delta_m$ for the low mass cut-off as described in \cite{LIGOScientific:2018jsj,Abbott:2020gyp}: see Appendix \ref{app:a} for the full expressions. The model PLG is thus completely determined by eight parameters. It is able to capture formation channels such as hierarchical formation in dense globular clusters. The Gaussian peak then represents a pile up of BBHs e.g., due to the PISN \cite{Abbott:2020gyp}.
Current data \cite{LIGOScientific:2018jsj,Abbott:2020gyp} suggests that  BBH formation is a mixture of the isolated and hierarchical formation channel and is thus better fitted by a PLG model. 

\subsection{Basics of the inference scheme}

We now present the general framework for joint population and cosmological inference. 

Given a set of $N_{\rm obs}$ GW detections associated with the data $\{x\}=(x_1,...,x_{\rm obs})$, the posterior on $\Lambda$ can be expressed as \cite{Mandel:2018mve,2019PASA...36...10T,Vitale:2020aaz}
\begin{equation}
    p(\Lambda|\{x\},N_{\rm obs}) \propto p(\{x\},N_{\rm obs}|\Lambda)p(\Lambda),
    \label{eq:gen}
\end{equation}
where $p(\Lambda)$ is a prior on the hyper-parameters. The term $p(\{x\},N_{\rm obs}|\Lambda)p(\Lambda)$ can be expanded as
\begin{equation}
    p(\{x\},N_{\rm obs}|\Lambda)p(\Lambda)=p(N_{\rm obs}|\Lambda) p(\{x\}|N_{\rm obs},\Lambda),
\end{equation}
where the term $p(N_{\rm obs}|\Lambda)$ is a Poisson distribution that relates the number of observed events $N_{\rm obs}$ with the expected number of detected events.
Since we are not interested in rate estimation in this work, we analytically marginalize over the total number of expected events by setting a scale-free prior \cite{Mandel:2018mve,Fishbach:2018edt}, which is also linked to the merger rates.

The term $p(\{x\}|N_{\rm obs},\Lambda)$ is the likelihood of observing the collection of the data $\{x\}$ given a set of population parameters and $N_{\rm obs}$ observed signals. If each of the signals is detected in a data chunk $x_i$, which is independent from the others, we can write
\begin{equation}
    p(\{x\}|N_{\rm obs},\Lambda) = \prod_i^{N_{\rm obs}} p(x_i|\mathscr{D},\Lambda),
\end{equation}
where $\mathscr{D}$ is the hypothesis (assumed true) of having a trigger from an astrophysical signal. (Below we will define triggers to be signals with SNR $\rho_{{\rm det},i}\geq 12$.)
%
The term $p(x_i|\mathscr{D},\Lambda)$ can be rewritten using Bayes theorem as
\begin{equation}
    p(x_i|\mathscr{D},\Lambda)=\frac{p(\mathscr{D}|x_i,\Lambda)p(x_i|\Lambda)}{p(\mathscr{D}|\Lambda)},
    \label{eq:singl}
\end{equation}
where $p(\mathscr{D}|x_i,\Lambda)$ is the probability of having a detection in the data $x_i$ and a set of cosmological parameters $\Lambda$. It is thus equal to 1 by assumption \cite{Mandel:2018mve}.
The likelihood of the GW event $p(x_i|\Lambda)$ given the population parameters can be factorized using the single source parameters $\theta$ as
\begin{equation}
    p(x_i|\Lambda) = \int p(x_i|\Lambda,\theta)p_{\rm pop}(\theta|\Lambda)d\theta,
    \label{eq:liksin}
\end{equation}
where $p_{\rm pop}(\theta|\Lambda)$ is the population-induced prior of Eq.~\eqref{eq:popinduced}.

The denominator $p(\mathscr{D}|\Lambda)$ in Eq.~(\ref{eq:singl}) is the probability of having a trigger of astrophysical origin, given a set of cosmological and population parameters. This is a normalization factor of the likelihood $p(x_i|\Lambda)$ and it describes what is usually referred to as \textit{selection effects} \cite{Mandel:2018mve,Vitale:2020aaz}.
This term can be written as an integral over every possible realization of detectors' data that will pass the detection threshold
\begin{equation}
    p(\mathscr{D}|\Lambda)=\int_{\rho_{{\rm det},i}\geq 12} \int p(x_i|\theta,\Lambda)p_{\rm pop}(\theta|\Lambda)d\theta dx_i.
\end{equation}
We assume that the noise properties are stationary, and hence the detectability of all events are the same meaning we can drop the subscript $i$.
The integral can then be written as \cite{Mandel:2018mve}
\begin{equation}
    p(\mathscr{D}|\Lambda)= \int p_{\rm det}(\theta,\Lambda)\:p_{\rm pop}(\theta|\Lambda)d\theta,
    \label{eq:selection}
\end{equation}
where $p_{\rm det}(\theta, \Lambda)$ is the probability of detecting the source with parameters $\theta$ and assuming the population and cosmological hyper-parameters $\Lambda$.
By substituting in Eq.~\eqref{eq:gen} the terms in Eqs.~\eqref{eq:liksin}-\eqref{eq:selection} we obtain the posterior on the source population and cosmological hyper-parameters
\begin{equation}
    p(\Lambda|\{x\},N_{\rm obs}) \propto p(\Lambda) \prod_{i}^{N_{\rm obs}} \frac{\int p(x_i|\Lambda,\theta)p_{\rm pop}(\theta|\Lambda)d\theta}{\int p_{\rm det}(\theta,\Lambda)p_{\rm pop}(\theta|\Lambda)d\theta}.
    \label{eq:posterior}
\end{equation}

\section{Application of the inference scheme to a simulated population of BBH}
\label{sec:catalogue1}

We now apply the above scheme to estimate cosmological and population parameters from a simulated population of BBH. In this section we present our population, and  first results on population and cosmological parameter inference. 

\subsection{Simulated BBH population}
\label{sec:populated}

We simulate a set of BBH GW events detected in LIGO and Virgo data assuming sensitivities comparable to the recent O2 and O3 observing runs \cite{TheLIGOScientific:2014jea,TheVirgo:2014hva,Acernese:2019sbr,2020arXiv200801301I}.

We choose a uniform in comoving volume merger rate $\gamma=0$, and draw the BBH component masses in the source frame from the PLG distribution. The power-law component is delimited by the two mass scales $\mmin=5 M_{\odot}$ and $\mmax=85 M_{\odot}$, and the slope for the primary mass distribution is $\alpha=2$, and of the mass ratio, $\beta=0$. We choose $\lambda_g=0.1$ so that $10\%$ of the total number of BBHs are in the Gaussian component. Its mass distribution has a mean $\mu_g=40 M_\odot$ and a standard deviation of $\sigma_g=5 M_\odot$. A tapering with a characteristic window of $\delta_m = 5 M_\odot$ is applied to the lower end of the distribution. The synthetic BBH catalog generated with this distribution is representative of the preferred model inferred from the GWTC-1 and GWTC-2 catalogs \cite{LIGOScientific:2018mvr,Abbott:2020niy}.

\begin{figure}[htp!]
    \centering
    \includegraphics[scale=0.4]{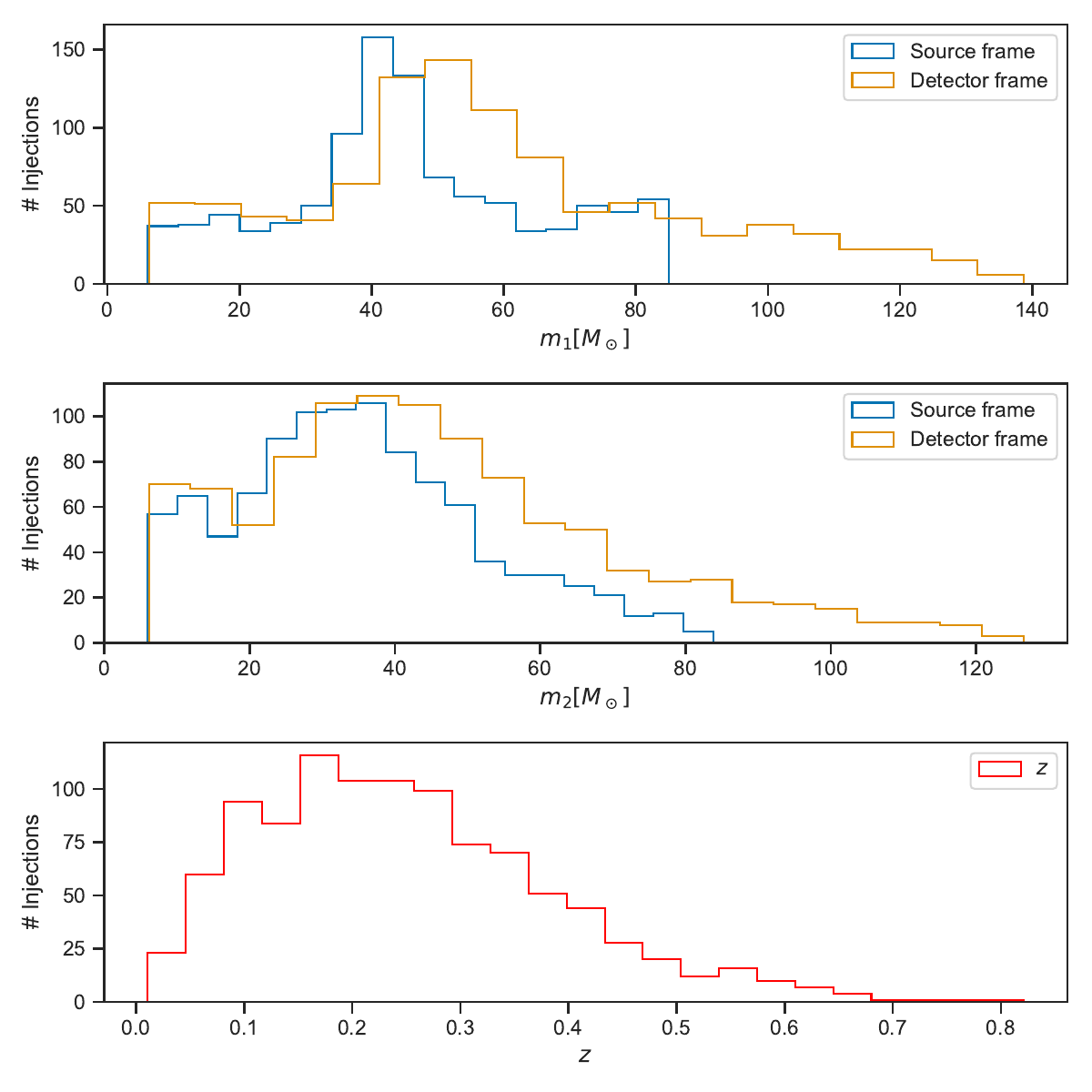}
    \caption{Simulated population of 1024 observed events, showing the mass distributions (in the detector and source frames) and redshift distribution.}
    \label{fig:nice}
\end{figure}

We choose $H_0=67.7 \hu$ and $\Omega_{m,0}=0.308$ \cite{Ade:2015xua}, and analyze $N_{\rm inj} \leq 1024$ simulated events that pass the SNR detection threshold $\rho_{\rm det} \geq 12$ \cite{Abbott:2019yzh}. The population is shown in Fig.~\ref{fig:nice}, and reaches a maximum redshift of $z\sim 0.8$.

For each simulated binary, we generate posterior samples for the masses and luminosity distance by following an approximation similar to \cite{Farr:2019twy}: all details may be found in Appendix \ref{app: b quick posterior samples}. A second study based on posterior samples produced by a proper ``end-to-end'' analysis (no approximation involved) is presented later in Sec.~\ref{sec:5}.

\begin{figure*}[!htp]
    \centering
    \includegraphics[scale=0.4]{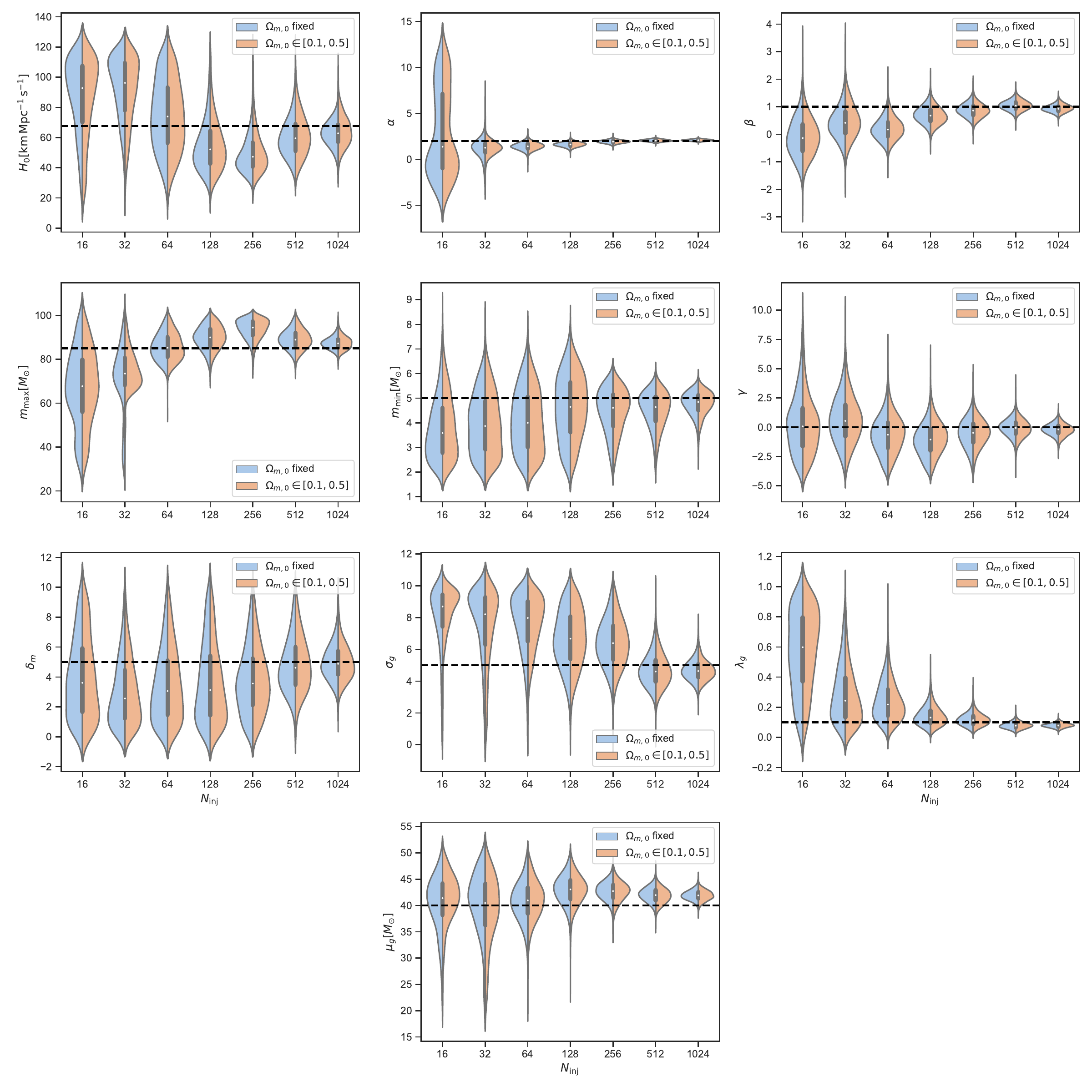}
    \caption{Posterior probability density distributions on the different population parameters (PLG mass model) as more and more GW detections are analyzed (horizontal axis). The horizontal black dashed line indicates the true parameters of the population. The blue posteriors are obtained by fixing $\Omega_{m,0}=0.308$, while the orange posteriors are marginalized over the estimation of $\Omega_{m,0}$.}
    \label{fig:convergence}
\end{figure*}

\subsection{Application of the inference scheme}

With this population, we now apply the inference scheme of section \ref{sec:2} to estimate jointly the hyperparameters, namely the mass model parameters, rate evolution $\gamma$, Hubble constant $H_0$ and mass-fraction $\Omega_{m,0}$. 

We consider two cases (i) $\Omega_{m,0}$ is fixed to the Planck value, $\Omega_{m,0}=0.308$ \cite{Ade:2015xua}, (ii) $\Omega_{m,0}$ is able to vary in the range $[0.1,0.5]$ with a uniform prior.

Figs.~\ref{fig:convergence} and \ref{fig:accuracies} show the marginal posterior distributions and the error on the population and cosmological parameters that we obtain as we analyse more GW events. All parameters are recovered to within $2\sigma$ of their true values. 
From Fig.~\ref{fig:convergence} we conclude that $\mmax, \mu_g$ and  $\alpha$ are the parameters that can be measured with the best accuracy, respectively at the 10\% and 8\% level and 11\%. The other population parameters can be measured within 30\% to 50\% accuracy with $\gtrsim 1000$ signals with the exception of the rate evolution parameter and the tapering factor. The rate evolution is the most difficult parameter to measure as we are looking at events at low redshift with current sensitivities.

The predicted  accuracy for $H_0$ is worse than that of \cite{Farr:2019twy} based on 5 years of observations for advanced LIGO. Two reasons explain this discrepancy: (\textit{i}) we consider sensitivities comparable to current detectors instead of future design sensitivities used in \cite{Farr:2019twy} and (\textit{ii}) our simulated population model leads to fewer detected BBH events ($\sim 15\%$ against to $25\%$ for \cite{Farr:2019twy}, see Fig.~\ref{fig:fraction_of_events}), thus reducing the events that are informative on the upper cut-off of the mass distributions, resulting in turn into a degraded $H_0$ estimation.

\subsection{Asymptotic normality and $1/\sqrt{n}$ error decay for large samples \label{sec:asymptotic}}

\begin{figure*}[htp!]
    \centering
    \includegraphics[scale=0.4]{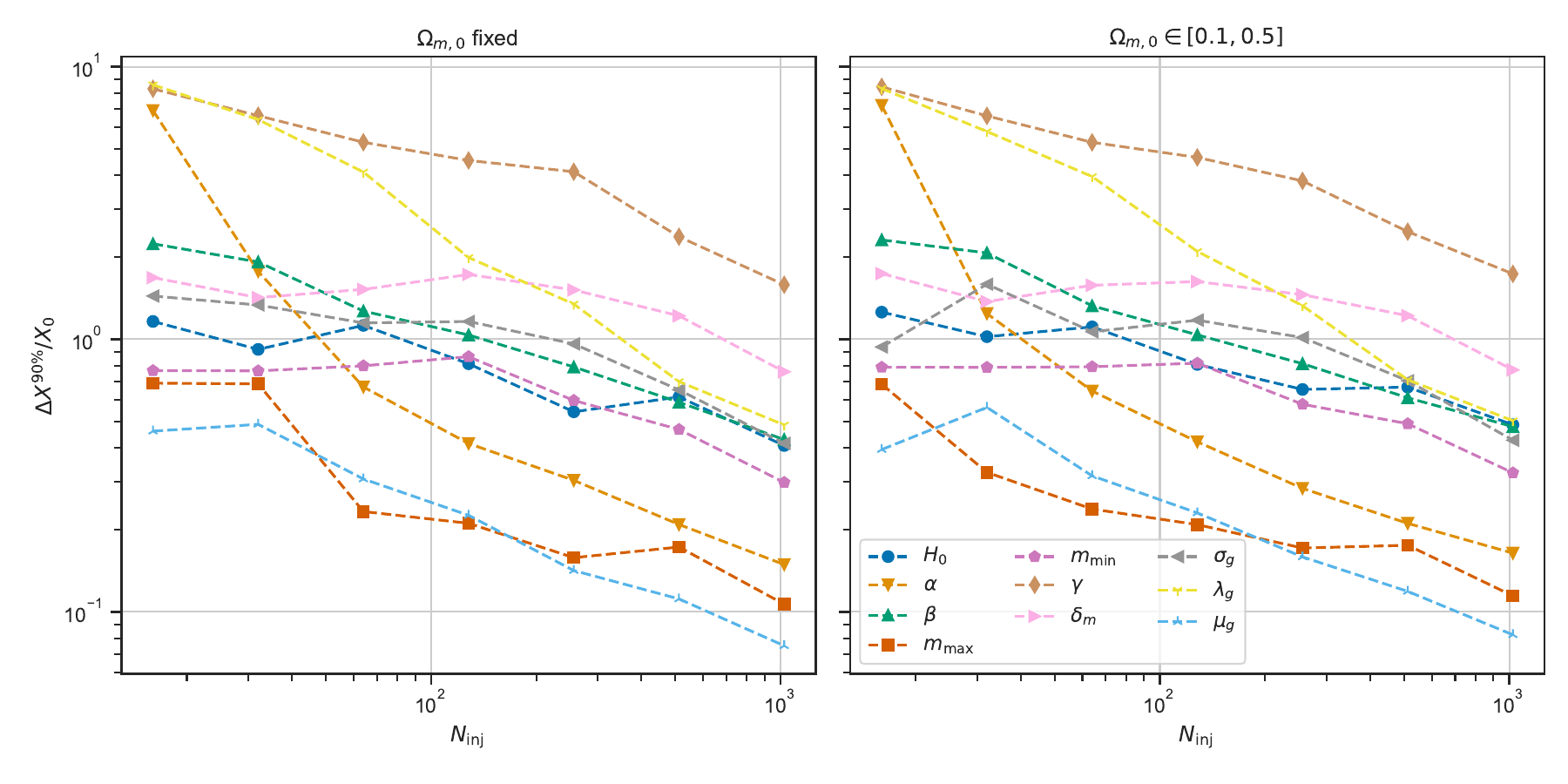}
    \caption{\textit{Left}: Accuracy at the 90\% CL on the population parameters (see levels) when we combine more and more GW detections. For this case we fix $\Omega_{m,0}=0.308$ to the injected value. \textit{Right}: Same but varying $\Omega_{m,0}$ in the range $\in [0.1,0.5]$.}
    \label{fig:accuracies}
\end{figure*}

The Bernstein-von Mises theorem (see e.g.~ \cite{van1998asymptotic}) states that, under mild assumptions (on the smoothness and continuity of the likelihood and prior distribution) and in the limit of large samples $n$, the posterior distribution tends to a normal distribution centered at the maximum likelihood estimate with standard deviation $\propto 1/\sqrt{n}$.
From Fig.~\ref{fig:convergence} we observe that the asymptotic regime is qualitatively reached when $N_{\rm{inj}} \gtrsim 500$ for most of the parameters (with the exception of $\mmin$ whose distribution remains skewed for large samples). Fig.~\ref{fig:accuracies} confirms these findings and shows a $1/\sqrt{n}$ error decay for all parameters in the limit of large $N_{\rm{inj}}$.

In the remainder of this paper we further discuss the results of this simulation, study their robustness with respect to initial priors, and also identify and quantify potential biases that may result from the interplay between hyper-parameters.

\section{Correlations between cosmological and mass-population parameters}
\label{sec:3}

In this section, we study the interplay between cosmological and mass-population parameters, focusing on some cases of particular relevance. We will show that amongst the parameters which have the strongest correlations are $(H_0,\mmax,\mu_g)$.  
In section \ref{sec:4} we will question what happens if we fix some of these parameters to incorrect values. 

\subsection{Weak impact of $\Omega_{m,0}$}
\label{sec:omegam}

\begin{figure}
    \centering
    \includegraphics[scale=0.5]{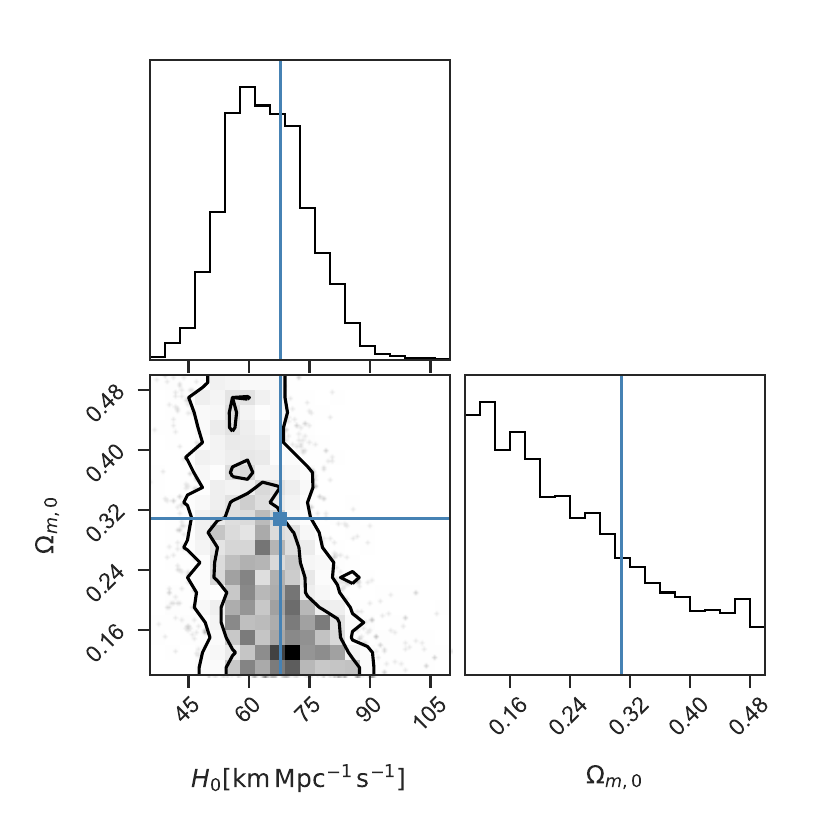}
    \caption{Posterior distribution on the $H_0$ and $\Omega_{m,0}$ for 1024 BBH events detected with LIGO and Virgo at current sensitivities. The blue lines show the true parameters.}
    \label{fig:H0_Om0}
\end{figure}

We find that $\Omega_{m,0}$ does not impact the estimation of the mass-related population parameters, see Figs.~\ref{fig:accuracies} and \ref{fig:H0_Om0}. It has weak impact on the estimation of $H_0$: 
in the specific case of our simulations, based on current detector sensitivities, this is observed above $\sim500$ detected events, when the accuracy on the $H_0$ estimation is of the order of $40\%$ (at 1.6$\sigma$~CL).

With $\sim$1000 GW detections, we estimate $H_0$ with a 40\% accuracy when fixing $\Omega_{m,0}$ to the true value, while this accuracy falls to $50\%$ if $\Omega_{m,0}$ is left to vary between $0.1$ and $0.5$.  This is due to the  correlation between $\Omega_{m,0}$ and $H_0$ in the GW luminosity distance, as can be seen in Fig.~\ref{fig:H0_Om0} which shows the marginal posterior distributions obtained with 1024 BBH events.

We conclude that with the current number of GW detections and sensitivities, one can neglect the unknown value of $\Omega_{m,0}$, but this should be reconsidered when analysing more GW events, especially if they are at higher redshifts.
This last comment is consistent with the conclusion of \cite{You:2020wju} for third generation detectors. 

In the remainder of this paper we set $\Omega_{m,0}=0.308$.

\subsection{Correlations between $H_0$ and features in the source-frame mass spectrum}
\label{sec:mmax}

Regarding the measurement of $H_0$, the most important parameters in the component mass spectrum are those that govern the high-mass features such as the maximum mass $\mmax$ and the position of the Gaussian peak $\mu_g$.

\begin{figure}
    \centering
    \includegraphics[scale=0.4]{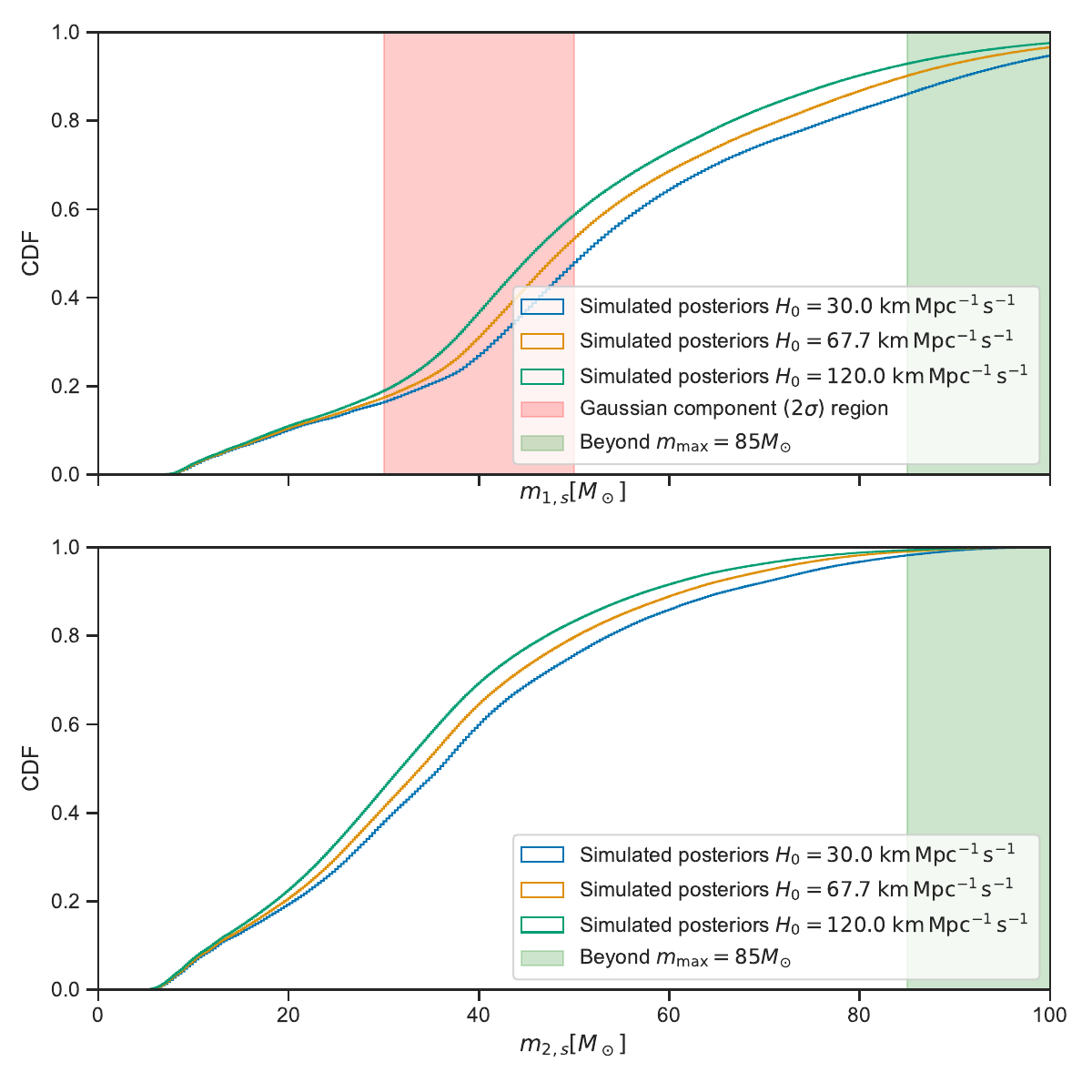}
    \caption{Cumulative posterior distribution of the source-frame masses (top: primary mass and bottom: secondary mass) inferred from the simulated population, and assuming different values for $H_0$. The position of the Gaussian peak and the maximum BH mass for BH are indicated by red and green areas respectively. See Sec.~\ref{sec:mmax} for an interpretation of this plot. The original prior has not been removed.}
    \label{fig:fraction_of_events}
\end{figure}

Fig.~\ref{fig:fraction_of_events} shows several cumulative posterior distributions for the source-frame masses, obtained by fixing $H_0$ to different values. For reference, the maximum BH mass $\mmax$ and the position of the Gaussian peak are indicated in the shaded areas. About 20\% to 40\% of the events have a primary mass $m_1$ estimate consistent with the position of the Gaussian component. Less than 20\% (and $\sim 10\%$ for $H_0 \sim 67 \hu$) of the events have a primary mass larger than $\mmax$. This decreases to a few percent for the secondary mass. These fractions set the scale for the number of events that carry information about the exact value for $\mmax$ and $\mu_g$. In addition, Fig.~\ref{fig:fraction_of_events} qualitatively explains the interrelation between these mass features and $H_0$. When $H_0$ varies between $30 \hu$ and $120\hu$ the above fractions of events that are informative on the two mass scales change by $\sim 20$\%, with natural consequences on the final accuracy for both the mass model parameters and $H_0$.

\begin{figure}
    \centering
    \includegraphics[scale=0.45]{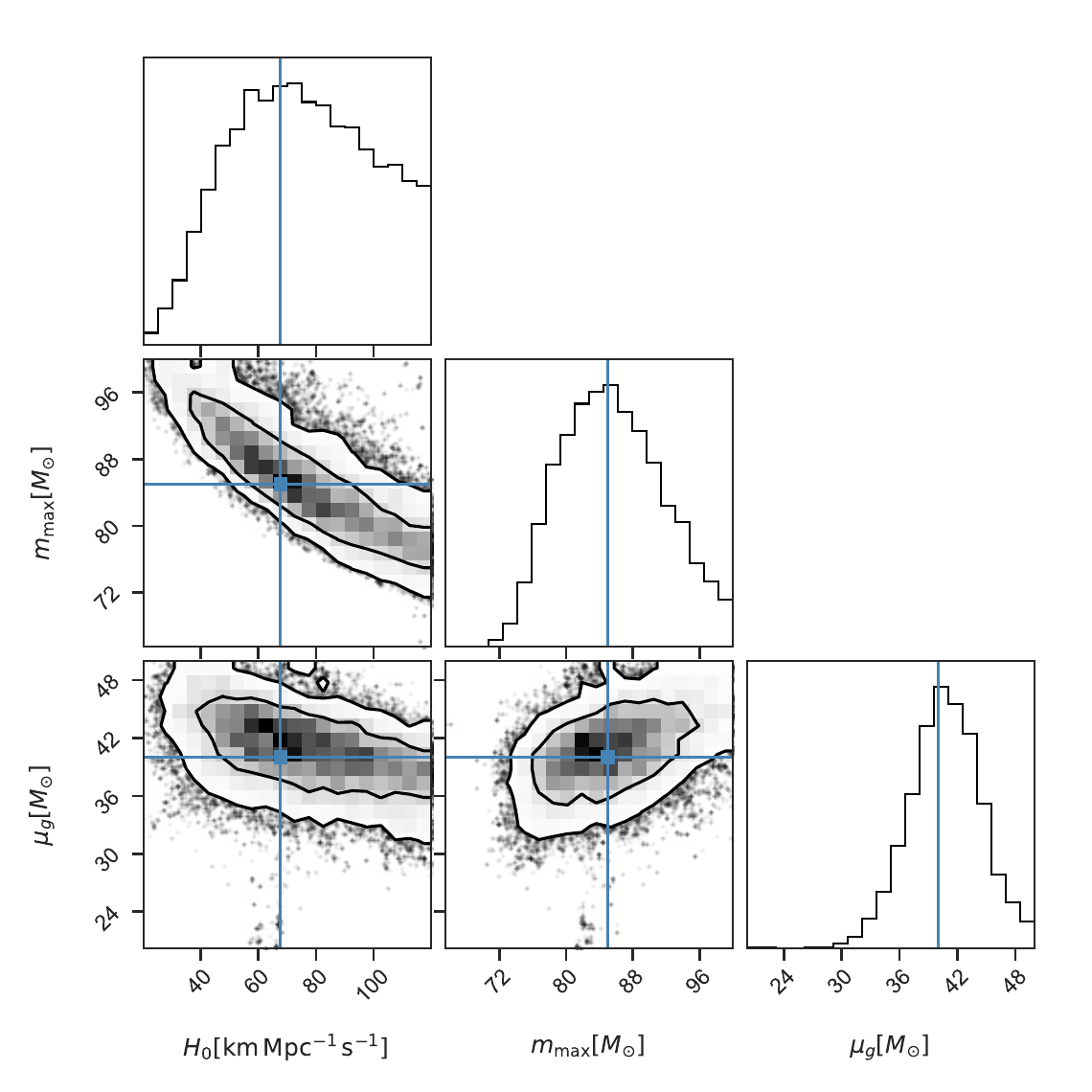}
    \caption{Posterior distribution on the $H_0$, $\mmax$ and $\mu_g$ for 64 BBH events detected with LIGO and Virgo at current sensitivities. The blue lines show the true parameters. The contours indicate the $1\sigma$ and $2\sigma$ confidence level intervals.}
    \label{fig:H0_mmax}
\end{figure}

The effect of the interplay between source-frame mass parameters and cosmology is clear in Fig.~\ref{fig:H0_mmax}. Considering 64 events (consistent with the current number of observed BBH by LIGO and Virgo), the joint $(H_0,\mmax,\mu_g)$ posterior distribution shows a strong correlation between the determination of $H_0$ and $\mmax$ and $\mu_g$.

In fact the determination of $\mmax$ and $\mu_g$ impacts the estimation of the $H_0$ in two ways. Concerning $\mmax$, first, lower $H_0$ values drag the observed GW source to lower redshifts, which in turn leads to higher source-frame masses. These are pushed towards $\mmax$: if they exceed this mass scale they become incompatible with the model. Therefore low $\mmax$ is incompatible with small $H_0$ values. Second, $\mmax$  also governs the fraction of detected events at higher masses. Since our model assumes masses up to $\mmax$, a lack of detected sources with masses close to the expected $\mmax$ should be compensated by lowering $\mmax$  or by decreasing $H_0$. 

Similar arguments are valid for the parameter $\mu_g$.  These two cross-correlations are clearly shown in Fig.~\ref{fig:H0_mmax} and play a r\^ole even when few events are observed.

While other parameters such as the rate evolution parameter might cause a bias in the estimation of $H_0$ (see Ref.~\cite{You:2020wju} for a discussion in the context of the Einstein Telescope), for current sensitivities 
$\mmax$ and $\mu_g$ (or any other equivalent parametrization of a sharp break in the observed mass spectrum) appear crucial for the inference of the cosmological parameters.

\section{Impact of population miscalibration on cosmological parameter estimation \label{sec:4}}

In this section we discuss the effect on the $H_0$ estimation of choosing a {\it different} mass model from that of the simulated population. The aim is to quantify the effect of possible population miscalibration.  


\subsection{Consequences of incorrect assumptions for the location of the mass features}

We have seen in Sec.~\ref{sec:mmax} that the parameters $\mmax$ and $\mu_g$ (or any other parameters related to features in the source-frame mass spectrum) play a fundamental r\^ole for the inference of $H_0$.  What is the consequence of fixing $\mmax$ and $\mu_g$ to a value inconsistent with their true values?

\begin{figure}
    \centering
    \includegraphics[scale=0.4]{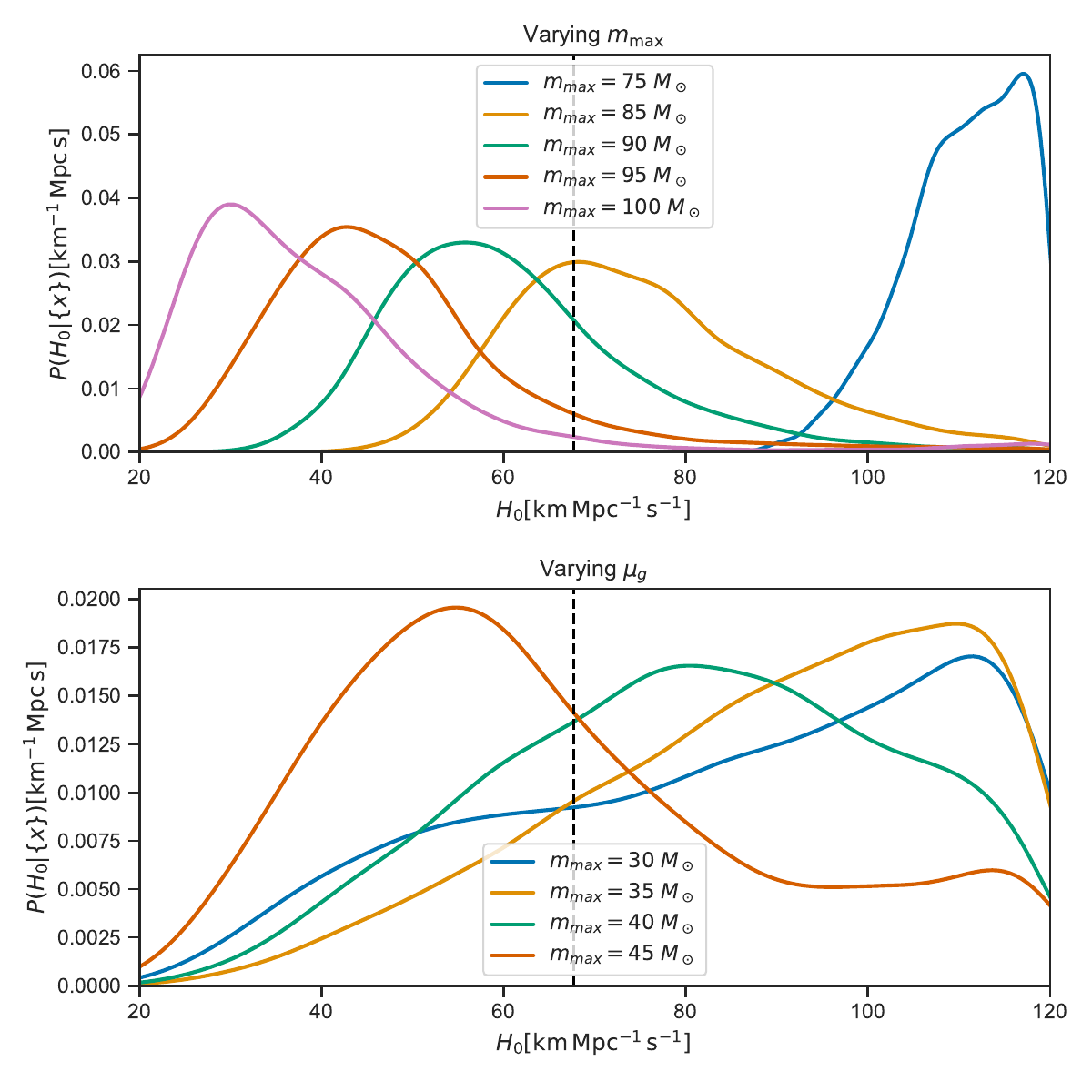}
    \caption{Posterior distribution for $H_0$ obtained by fixing $\mmax$ and $\mu_g$ in a range around their true values  $ \mmax = 85 M_{\odot}$ and $\mu_g = 40 M_{\odot}$. The black dashed line indicates the true value of $H_0$.}
    \label{fig:bias_H0}
\end{figure}

Fig.~\ref{fig:bias_H0} shows the marginal posterior distribution obtained for $H_0$ when fixing either $\mu_g$ or $\mmax$ to a wrong value and marginalizing over the rest of the population parameters. 
This figure is computed with 64 GW events, and is thus representative of the analyses that can be done with the current number of observed events in the GWTC-1 and GWTC-2 catalogs.
We observe that $H_0$ is biased toward smaller values when either $\mmax$ or $\mu_g$ are much higher than their true values.  
Conversely, when they are set too \textit{low}, $H_0$ is biased towards higher values.

In summary fixing the maximum mass for BH production can thus lead to biased estimations of the cosmological and source population parameters and in particular of $H_0$.

\subsection{Consequences of using an incomplete model}

We now discuss the impact of selecting an incomplete population model that misses some of the features of the real underlying mass spectrum (in our case the PLG model, with parameters specified in Section \ref{sec:populated}: a Gaussian peak at $\mu_g=40 M_\odot$ (with a standard deviation of $5 M_\odot$) and $\mmax = 85 M_{\odot}$). 
In particular, we study the recovery of the population parameters when we fit a PL model that thus misses the Gaussian peak component and tapering in the low-mass range.  We compare this with the full analysis (namely using the correct PLG population model).

\begin{figure}
    \centering
    \includegraphics[scale=0.45]{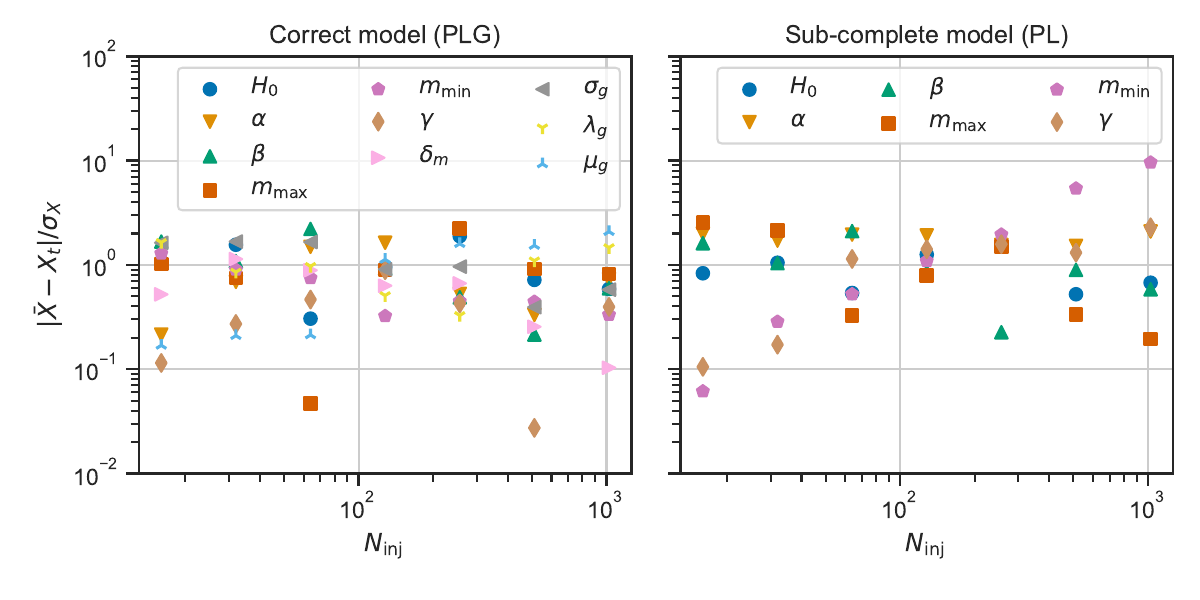}
    \caption{Discrepancy in terms of $\sigma$ between the estimated and true value of parameter $X$ (where $X$ is any of the parameters listed in the legend). The true population model is a PLG. \textit{Left plot}: Fit the correct model (PLG) ($\Omega_{m,0}$ fixed). \textit{Right plot}: Fit of an incorrect and incomplete model (PL).}
    \label{fig:bias_PLG.pdf}
\end{figure}

First we fit a PL model to the data. Fig.~\ref{fig:bias_PLG.pdf} shows the discrepancy (in terms of number of $\sigma$) between the estimated and true values for the population parameters. For low numbers of GW detections (low-sample regime) this figure of merit may not be very robust as posteriors may have tails. However, for large number of events the posteriors ``gaussianize'', and we should find the true values in a reasonable confidence interval $\sim 2 \sigma$.  The estimation of $\mmin$ departs from the true value by more than 10 $\sigma$.
The reason for failing to estimate $\mmin$ correctly is the lack of tapering at low masses for the PL model. The PL model is able to recover the correct value of $\mmax$ because, in the underlying population model, the separation in scales between the Gaussian component and $\mmax$ is more than 5$\sigma$.  Similarly it recovers the correct value of $H_0$.

Using a PLG model to fit the population, we obtain that for any number of detected GW signals, the true population parameters are within the $2 \sigma$ confidence levels. 

\begin{figure*}[htp!]
    \centering
    \includegraphics[scale=0.5]{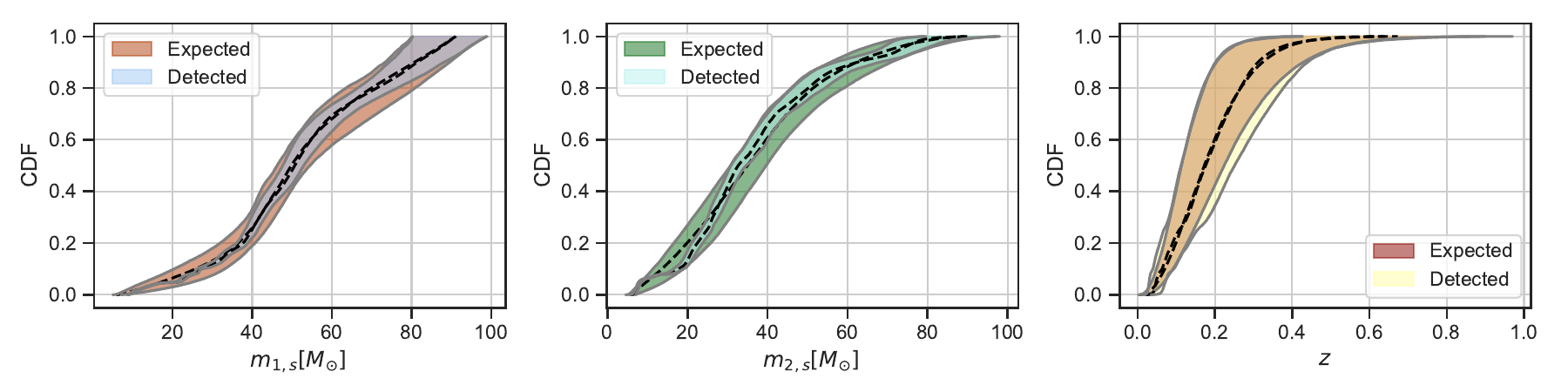}
    \includegraphics[scale=0.5]{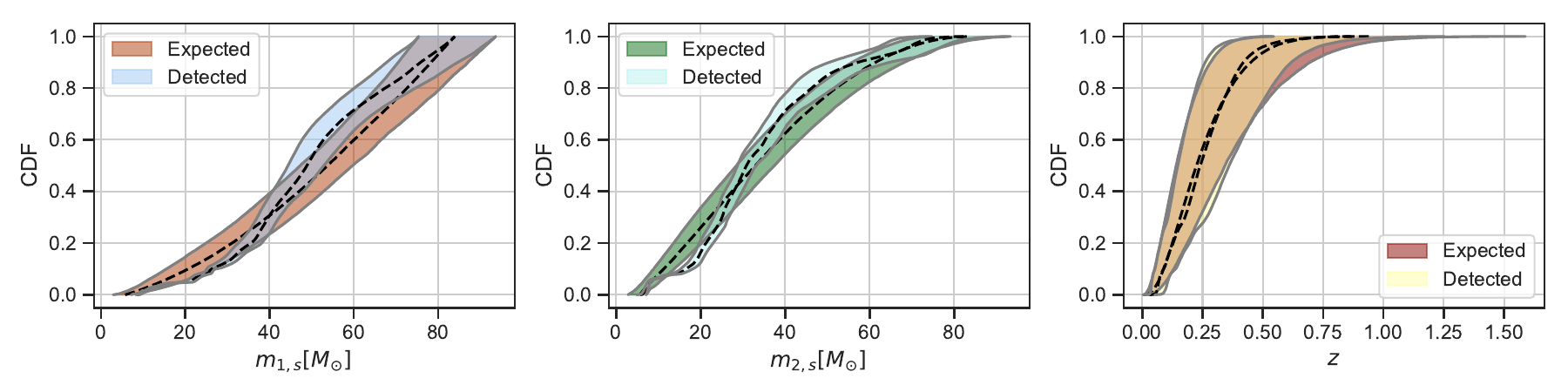}
    \caption{\textit{Top}: Posterior predictive check labelled as ``expected'' for the two source-frame masses and redshift fitting an underlying PLG population with the correct mass model in comparison with 64 observed events, labelled as ``detected''. The detected distributions match the posterior predictive checks. \textit{Bottom}: Same but fitting a sub-complete PL model to the underlying PLG model. The model struggles to fit the lower end part of the mass distribution and the excess of BHs around 40-50 $M_\odot$.}
    \label{fig:ppc_64}
\end{figure*}

While the estimation of the parameters common to the PL and PLG model are broadly consistent (except for $\mmin$), the PL model actually leads to an inaccurate fit of the observed population (it misses the Gaussian peak). This can be observed by calculating the Bayes factors as done in \cite{Abbott:2020gyp}, and also through posterior predictive checks as presented in Fig.~\ref{fig:ppc_64}.
This check consists in overlapping the \textit{expected} distribution of GW detections, obtained using the estimated population parameters, namely:
\begin{align}
    p_{\rm pop}(\theta|\{x\},N_{\rm obs})&= \int p_{\rm det}(\theta,\Lambda) \: p_{\rm pop}(\theta|\Lambda) \times \nonumber \\ &  p(\Lambda|\{x\},N_{\rm obs}) \:d\Lambda
\end{align}
with the distribution of \textit{detected} events (using population-induced priors). If the model is correct, the two cumulative distributions agree and the sanity check is passed.

The test presented in Fig.~\ref{fig:ppc_64} is computed with 64 BBHs events. While the incomplete PL model is able to infer the maximum mass of the underlying distribution, it fails to accommodate the lower end of the tapered mass distribution of the PLG model and the excess of BBHs between 40 and 50 $M_\odot$.

In conclusion, (over)simplified population models must be handled with care as this may lead to significant bias when the true mass spectrum has a complex shape. It is therefore essential to make a thorough goodness-of-fit evaluation of several models using Bayes factors. In the above example this would have shown that PLG was the preferred model.

\section{End-to-end analysis from gravitational-wave data}
\label{sec:5}

The results presented in Sec.~\ref{sec:3} and \ref{sec:4} are based on an approximated likelihood (see Appendix \ref{app: b quick posterior samples}), which allowed for fast generation of posterior distributions for large numbers of GW events. In this section we validate our results by an ``end-to-end'' analysis using {\it simulated GW $h(t)$ data and posterior samples generated from Bayesian samplers} used for GW parameter estimation. This analysis thus mirrors the analysis chain used for observational data  for the first time in literature.
In the following we present results of a joint cosmological and source population inference using the expected sensitivity for the future observing run O4.

We again simulate a mock BBH catalog, and now generate the associated GW signals using the \texttt{IMRPhenomD} \cite{2016PhRvD..93d4006H} waveform approximant. We retain 100,000 signals detected with $\rho_{\rm det} \geq 12$ by the LIGO-Virgo three-detector network at design sensitivity\footnote{\url{https://dcc.ligo.org/LIGO-T2000012/public}}.
From this catalog we select 200 events mimicking a population with a PL mass distribution with parameters $\alpha=2$, $\beta=0$, $\mmin=35 ~M_\odot$ and $\mmax=65~M_\odot$. 
The choice for $\mmin$ is not representative of realistic astrophysical expectations: it is made to speed up the analysis by avoiding the Bayesian estimation of low-mass events which takes substantially more time.
We furthermore fix the merger rate parameter $\gamma=2$, and the injected sources are taken to lie in the range $0\leq z \leq 2$.  The spins are assumed aligned with the orbital momentum. 
 
The selected subset of 200 events is processed by the inference pipeline Bilby using the Bayesian sampler \texttt{dynesty} \cite{2019ApJS..241...27A}. We run a full 10-dimensional parameter estimation (since we fix the coalescence time of the merger and assume aligned spins). We assume standard priors on the spin amplitudes, the polarization angle, sky position, inspiral phase, a $d_L^2$ prior on luminosity distance (which is later removed in the population analysis) and flat priors on the detector frame masses. For this latter sampling we do not impose the $m_{2,d}<m_{1,d}$ condition on the component masses, but apply it \textit{a posteriori}.

\begin{figure*}
    \centering
    \includegraphics[scale=0.5]{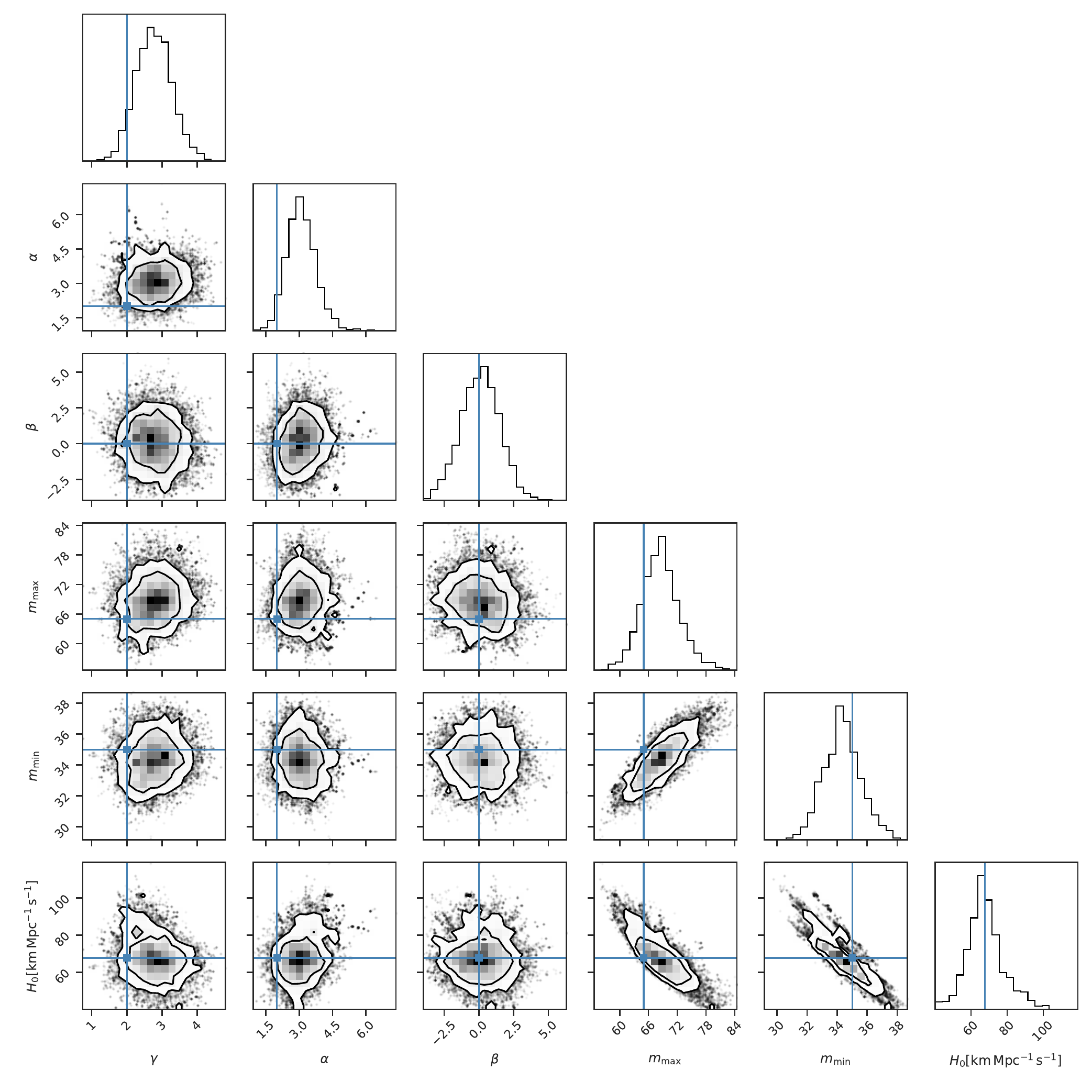}
    \caption{Posterior probability density distributions for the population parameters and the $H_0$ using posterior from 200 events generated with full parameter estimation. The blue lines indicate the population injected values. Levels indicate the 68\% and 90\% confidence intervals.}
    \label{fig:bilby_200_ev}
\end{figure*}

We find that the Bilby runs produce, for 90\% of the simulated events (symmetric intervals around the median values), uncertainties at the 90\% confidence level on luminosity distance, primary and secondary masses (detector frame) which are between 50\%-94\%, 17\%-32\% and 29\%-70\% respectively. (For comparison, for the same population, the likelihood approximant implemented in the previous sections would have given 26\%-94\%, 10\%-46\% and 10\%-40\% uncertainties on luminosity distance and detector frame masses for the same kind of population. As we can see, the likelihood approximant predicts a lower error budget on the secondary mass component.)
We then perform the analysis outlined in Sec.~\ref{sec:2} to estimate the population parameters, jointly with $H_0$. Fig.~\ref{fig:bilby_200_ev} shows the marginal posterior distributions obtained with the 200 selected events. 

The posterior distributions for the cosmological and population parameters are in good agreement with the true value. This thus provides a proof-of-principle for the applicability of the approach to real data. 
Though hundreds of events with non-Gaussian individual posteriors are combined, the posterior distributions converge to a normal distribution as noted in Sec~\ref{sec:asymptotic}. 

We notice a significant correlation between the lower-mass limit $\mmin$ and $H_0$, that was {\it not} present in the earlier simulations. This is simply a consequence of the higher value for $\mmin$ used here.  A much larger number of events is now informative on the lower mass cut-off of the mass spectrum, which can thus be accurately measured. Together with $\mmax$ the measurement of $\mmin$ provides an additional well-defined mass scale that correlates with $H_0$. This does not impact the final accuracy level of the $H_0$ measurement which appears the same as in the case when $\mmin=5 M_{\odot}$.

\section{Impact of population assumptions on the cosmology inferred from the O2/GWTC-1 catalog}
\label{sec:reanalysis}

In this section we re-examine the estimation of $H_0$ obtained with GWTC-1 in \cite{Abbott:2019yzh} in light of the above observations showing how population assumptions impact the $H_0$ result.

The analysis in \cite{Abbott:2019yzh} is based on the ``brightest'' BBHs of the GWTC-1 catalog selected with $\mathrm{SNR} > 12$ (6 events in total). The $H_0$ measurement uses redshift information from the GLADE \cite{Dalya:2018cnd} and DES \cite{Abbott:2018jhe} galaxy catalogs. Out of the six considered BBHs events, two have a low probability for their hosting galaxy to be in the galaxy catalogs (GW170104 and GW170809), three have a medium probability (GW150914, GW151226 and GW170608), while one has a probability almost equal to 1 (GW170814). In the limit of empty galaxy catalog (0\% completeness), the galaxy catalog analysis collapses to the analysis framework presented in Sec.~\ref{sec:2} and thus information on cosmological parameters might come from source-frame mass assumptions.

The analysis in \cite{Abbott:2019yzh} needs to \textit{a priori} assume a source population model, and the model chosen there is the PL model. It is based on \text{GWCOSMO} \cite{Gray:2019ksv} fixing $\Lambda_m =\{\mmax= 100~M_{\odot},\: \mmin=5~M_{\odot},\: \alpha=1.6,\: \beta=0\}$. These values were chosen to accommodate all possible values for the source-frame masses of the GW events in the GWTC-1 and GWTC-2 catalogs for any choice of $H_0 \in [20,140] \hu$.
Based on those assumptions the analysis draws samples from the posterior
\begin{equation}
    p(H_0|\Lambda_{m},x)= p(x|H_0,\Lambda_m)\:p(H_0).
    \label{eq:gwcosmo}
\end{equation}

The search in \cite{Abbott:2019yzh} partially explores the systematics introduced by the choice of PL model, trying several values of $\mmax$ and $\alpha$. No clear criteria are presented to guide the choice of the population parameters, though it was shown that certain choices effect the $H_0$ estimation.

In this section, we show that the inference scheme presented in Sec.~\ref{sec:2} can be used to robustly predict exactly how population parameters contribute to the $H_0$ estimation with a galaxy catalog analysis.

We run the joint population and cosmology analysis using the same set of GWTC-1 BBH events. We fix $\beta$ and $\gamma=0$ and allow $\mmin$, $\mmax$ and $\alpha$ to vary. We specifically target the region associated with the current tension on the $H_0$ estimate, and use for $H_0$ a uniform prior in the range $[67,\:74] \hu$. With these settings, the maximum likelihood is reached at $H_0=69 \hu$ with the parameters $\mmin=8.6 M_{\odot}$, $\mmax=37.5 M_{\odot}$ and $\alpha=2.2$. Those parameters best fit the data in the region of interest for $H_0$ but may not for other values. 

\begin{figure}
    \centering
    \includegraphics[scale=0.2]{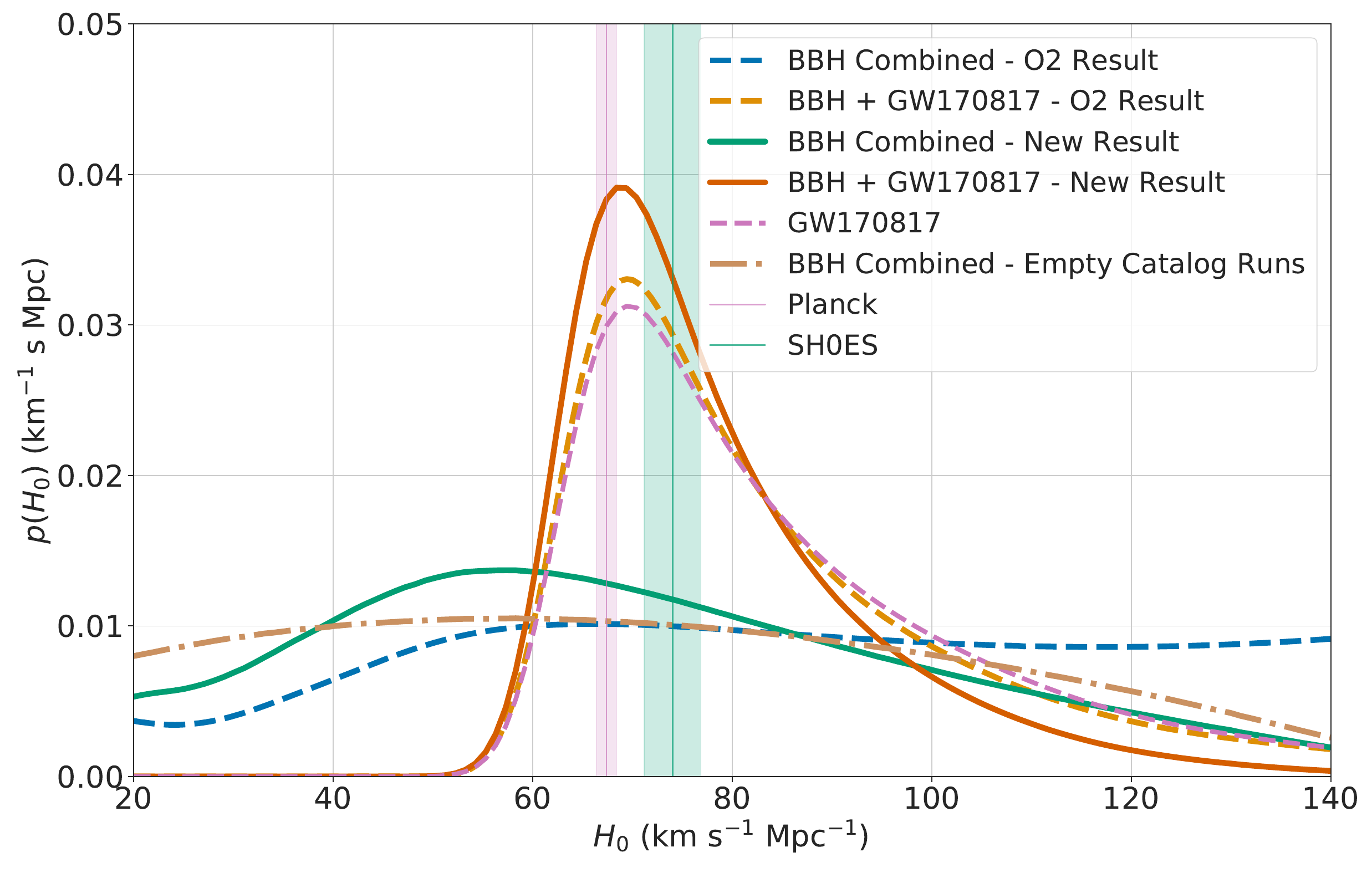}
    \caption{Posterior distribution on $H_0$ using the 6 GWTC-1 events with $\mathrm{SNR}>12$ and the GLADE and DES galaxy catalogs. The plot compares the results obtained in \cite{Abbott:2019yzh} with the new results of this paper (see discussion in Sec.~\ref{sec:reanalysis}).}
    \label{fig:results_vs_O2}
\end{figure}

In a second step we apply the analysis of \cite{Abbott:2019yzh} using the GLADE and DES galaxy catalogs using the new set of population parameters. Fig.~\ref{fig:results_vs_O2} shows the results for both approaches. We obtain the credible interval $H_0 = 68^{+13}_{-7} \hu$ to be compared to $H_0 = 68^{+16}_{-8} \hu$ reported in \cite{Abbott:2019yzh}. The width of the former is about $15 \%$ narrower; the $H_0$ estimate is thus more informative in the tension region. 
In Fig.~\ref{fig:results_vs_O2} the posterior tails appear considerably reduced with the new choice of population parameters; this is not surprising, as the population parameters are chosen to maximize the likelihood in the central $H_0$-tension region.

The analysis with galaxy catalogs entails the joint marginalization over both the cosmological and population parameters. If it is impossible to marginalize because of computational limitations as explained in App.~\ref{app:d}, the population analysis presented above allows to quantify the potential impact of a specific choice of population.

This case study shows population assumptions matter as they impact on the final measurement accuracy. In the absence of a strong prior belief for the population model, this advocates for analysis schemes that \textit{consider population and cosmological parameters jointly} and not separately. This suggests to perform joint source population and cosmological inference together with the use of galaxy catalogs. Combining the two analyses is not obvious and likely leads to challenging computational issues. If this turns out to be intractable, a comprehensive evaluation of the systematics induced by population assumptions are required to deduce robust conclusions from analyses that treat source population and cosmology separately.

\section{Impact of the population parameters on $H_0$ when an EM counterpart is observed}
\label{sec:EM}

We end this paper by considering a  different situation: namely we now {\it suppose  that an EM counterpart is detected in association with each GW event in the population}. We assume this will give an independent redshift measurement $z_{\rm obs}$ for each event, as for GW170817, and in this section we consider the impact of this additional data on the estimation of $H_0$.
We will show (modulo some caveats, see later) that when an EM counterpart is observed, the choice of population parameters does not impact the $H_0$ estimation.

In this case, the hierarchical posterior in Eq.~(\ref{eq:posterior})
is modified to account for the additional data, leading to (we drop the subscript $i$):
\begin{align}
     p(\Lambda|x,z_{\rm obs}) & \propto p(\Lambda) \: p(x,z_{\rm obs}|\Lambda) \nonumber \\ \propto p(\Lambda) & \frac{\int p(x|\Lambda,\theta) p (z_{\rm obs}|z,\bar{\theta}) p_{\rm pop}(\theta|\Lambda)d\theta}{\int p_{\rm det,GW}(\theta,\Lambda) p_{\rm det,EM}(\theta,\Lambda)p_{\rm pop}(\theta|\Lambda)d\theta},
    \label{eq:EMcount}
\end{align}
where we have separated the source redshift $z$ from the other binary parameters, writing $\theta = \{z,\bar{\theta}\}$. The term $p(z_{\rm obs}|z,\bar{\theta})$ is the likelihood of measuring a redshift $z_{\rm obs}$ given the true source redshift $z$ and other binary parameters $\bar{\theta}$.
Finally the selection effects connected to EM observations are taken into account through $p_{\rm det,EM}(\theta,\Lambda)$.

Eq.~(\ref{eq:EMcount}) can be simplified under the following assumptions (i) the redshift measurement is very accurate and independent of the binary parameters, i.e.~$p(z_{\rm obs}|z,\bar{\theta})\approx \delta(z_{\rm obs}-z)$; (ii) measurement of the luminosity distance $d_L$ and detector frame masses are mutually independent, i.e.~$p(x|d_L,m_{1,d},m_{2,d}) \propto p(x|d_L) p(x|m_{1,d},m_{2,d})$. Then Eq.~\eqref{eq:EMcount} simplifies to (see Appendix \ref{app:scream})
\begin{align}
     p(\Lambda|\{x\},z_{\rm obs}) & \propto \frac{p(\Lambda)}{[ p(\mathscr{D}|\Lambda)]^{N_{\rm obs}}} \nonumber \\ &\times \prod_i^{N_{\rm obs}}  p(z^{i}_{\rm obs}|\Lambda_c) p(x^i|d_L(\Lambda_c,z^{i}_{\rm obs}))
     \nonumber  \\
     & \times \prod_i^{N_{\rm obs}}  I(x^{i};\Lambda_m,z^{i}_{\rm obs}),
     \label{eq:EM_simp_ter}
\end{align}
where $I$ is defined in Eq.~(\ref{eq:Idef}) and
\begin{equation}
    p(\mathscr{D}|\Lambda)= \int p_{\rm det,GW}(\theta,\Lambda)p_{\rm det,EM}(\theta,\Lambda)\:p_{\rm pop}(\theta|\Lambda)d\theta.
\end{equation}

It is important to notice that the two last terms depend individually on either the population or cosmological parameters. 

Fixing the population parameters $\Lambda_m$ to incorrect values thus results in a biased evaluation of the last line of Eq.~(\ref{eq:EM_simp_ter}). This term enters in the inference of $\Lambda_c$ simply as a normalization constant and thus does not lead to any bias. {\it If} the term $p(\mathscr{D}|\Lambda)$ that accounts for selection effects is close to a separable function of $\Lambda_m$ and $\Lambda_c$, then fixing an incorrect value of $\Lambda_m$ will not bias $\Lambda_c$. Physically we expect this to be the case for current ground based detectors (the selection bias is not introduced by the redshifting of the source-frame masses outside their sensitive mass range).

\begin{figure}
    \centering
    \includegraphics[scale=0.5]{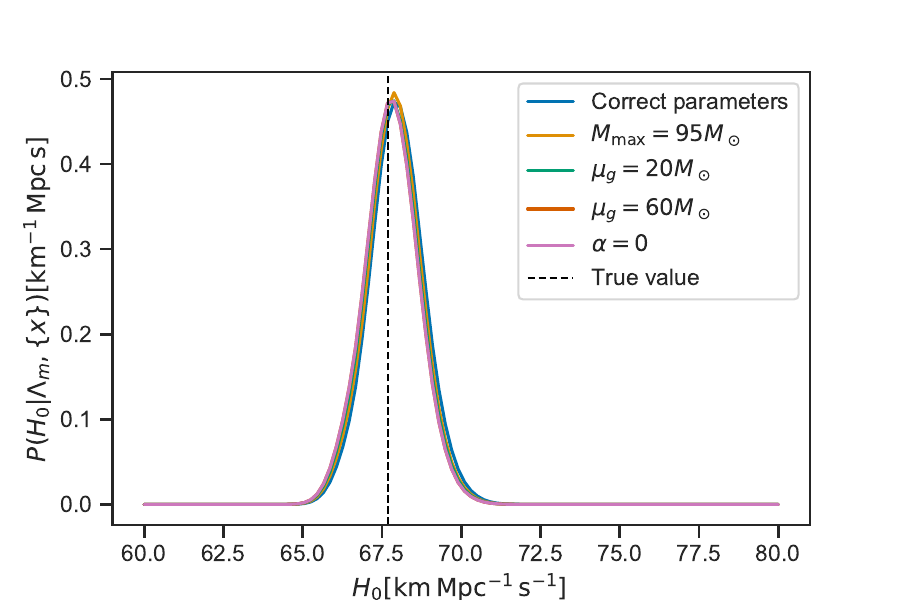}
    \caption{Hubble constant posterior generated from 64 our synthetic population of BBHs (section \ref{sec:populated}) fixing different population models and providing the redshift of the GW source (assumed from an EM counterpart). A incorrect choice of one of the population parameters (see legend) does not affect significantly the $H_0$ estimation. The vertical dashed line indicate the injected value. Note that for this plot the posterior samples are generated taking into account the correlations between masses and luminosity distance (point (ii) above is dropped).}
    \label{fig:em_counterpart}
\end{figure}

We conclude that in presence of an EM counterpart, incorrect population priors do not affect the cosmological parameters. This is confirmed by Fig.~\ref{fig:em_counterpart} where the posterior on $H_0$ computed with 64 events of our synthetic population of Sec.~\ref{sec:populated} is shown for different population models.

We end with a final precautionary note: even though the $H_0$ posterior is weakly sensitive to the population parameters in this case, the population parameters can be {\it totally} mismatched --- for instance the source-frame mass model should include the source-frame masses we observe (otherwise one of the $I$ functions in Eq.~\eqref{eq:EM_simp_ter} could vanish). 

\section{Conclusions \label{sec:6}}
\label{sec:conc}

In this paper we have discussed the impact of population assumptions for cosmological inference with GW events. We have shown that, even with current sensitivities, population assumptions on the features of the mass spectrum can affect the estimation of the cosmological parameters $H_0$ and $\Omega_{m,0}$.

We have shown that the parameters that govern the position of the middle peak or of high-mass cut-off in the mass spectrum are strongly correlated with the final estimated value for $H_0$. We have also shown that incorrect priors on the properties of those features introduce a significant bias. 

Together with the uncertainties of GW data calibration \cite{2020arXiv200910192V}
and the misevaluation of selection effects due to the BNS viewing angle \cite{2020PhRvL.125t1301C}, population assumptions could represent the major and possibly dominant source of systematics for GW-based cosmology with current and future GW observations.  That is why we argue that cosmological and population parameters should be performed jointly. However, this can be computationally challenging given the large number of galaxies that has to be considered and the rapidly increasing number of GW detections. Those computational challenges can possibly be resolved by porting the inference code to GPU \cite{Talbot:2019okv}.

We conclude that GW-based cosmological analysis should be complemented with a comprehensive and in-depth evaluation of the impact of population assumptions on the final result.

\vspace{3mm}
\paragraph*{Acknowledgments} --  SM is supported by the LabEx UnivEarthS (ANR-10-LABX-0023 and ANR-18-IDEX-0001), of the European Gravitational Observatory and of the Paris Center for Cosmological Physics. KL is grateful to the Fondation CFM pour la Recherche in France for supporting his PhD.
The authors are grateful for computational resources provided by the LIGO Laboratory and supported by National Science Foundation Grants PHY-0757058 and PHY-0823459. CK is partially  supported   by  the Spanish MINECO   under the grants SEV-2016-0588 and PGC2018-101858-B-I00, some of which include ERDF  funds  from  the  European  Union. IFAE  is  partially funded by the CERCA program of the Generalitat de Catalunya. RG is supported by the Science and Technology Facilities Council. SMu is supported by the Delta ITP consortium, a program of the Netherlands Organisation for Scientific Research (NWO) that is funded by the Dutch Ministry of Education, Culture, and Science (OCW). LIGO is funded by the U.S. National Science Foundation. Virgo is funded by the French Centre National de Recherche Scientifique (CNRS), the Italian Istituto Nazionale della Fisica Nucleare (INFN), and the Dutch Nikhef, with contributions by Polish and Hungarian institutes. This material is based upon work supported by NSF’s LIGO Laboratory which is a major facility fully funded by the National Science Foundation.

\newpage
\clearpage

\appendix

\onecolumngrid 

\section{Mass models \label{app:a}}

In this appendix we provide the mathematical expressions of the two phenomenological mass models used in this paper. These are the same employed in \citep{Abbott:2020gyp}. 
The two mass models combines two baseline probability density functions. The first is the power-law distribution with slope $\alpha$, truncated to the interval defined by the lower and upper bounds $x_{\rm min}$ and $x_{\rm max}$:
\begin{equation}
\mathcal{P}(x|x_{\rm min},x_{\rm max},\alpha) \propto 
\begin{cases}
    x^\alpha & \text{for $x_{\rm min}\leq x \leq x_{\rm max}$} \\
    0 & \text{otherwise}.
\end{cases}
\end{equation}

The second is the Gaussian distribution with mean $\mu$ and standard deviation $\sigma$,
\begin{equation}
\mathcal{G}(x|\mu,\sigma)=\frac{1}{\sigma\sqrt{2\pi}} e^{-\frac{(x-\mu)^2}{2\sigma^2}}.
\end{equation}

The source-frame mass priors for the BBHs population are factorized as 
\begin{equation}
\pi(m_{1,s},m_{2,s}|\Lambda_m)=\pi(m_{1,s}|\Lambda_m)\pi(m_{2,s}|m_{1,s},\Lambda_m),
\end{equation}
where $\pi(m_{1,s}|\Lambda)$ is the distribution of the primary mass component while $\pi(m_{2,s}|m_{1,s},\Lambda)$ is the distribution of the secondary mass component given the first.

We consider the following two mass distributions:

\vspace{3mm} 
\noindent \textbf{Power-law model (PL)}:  this model defines the distribution of the primary mass $m_{1,s}$ as a truncated power-law with slope $-\alpha$ between the minimum mass $\mmin$ and the maximum mass $\mmax$, namely
\begin{equation}
    p(m_{1,s}|\mmin,\mmax,\alpha)=\mathcal{P}(m_{1,s}|\mmin,\mmax,-\alpha).
\end{equation}

\vspace{3mm}
\noindent \textbf{Power-law model with Gaussian component (PLG)}: this model defines the primary mass component as a superposition of a truncated power-law  with slope $-\alpha$ between the minimum mass $\mmin$ and the maximum mass $\mmax$ plus a Gaussian component with mean $\mu_g$ and standard deviation $\sigma_g$, namely
\begin{eqnarray}
    p(m_{1,s}|\mmin,\mmax,\alpha,\lambda_g,\mu_g,\sigma_g)=(1-\lambda_g) \mathcal{P}(m_{1,s}|m_{\rm min},m_{\rm max},-\alpha)+\lambda_g \mathcal{G}(m_{1,s}|\mu_g,\sigma).
\end{eqnarray}
We also apply a smoothing factor to the lower end of the mass distribution
\begin{equation}
p(m_{1,s},m_{2,s}|\Lambda_m)=p(m_{1,s}|\Lambda_m) S(m_1|\mmin, \delta_m) \: p(m_{2,s}|m_{1,s},\Lambda_m)S(m_2|\mmin, \delta_m),
\end{equation}
where $S$ is a sigmoid-like window function that performs a tapering of the lower end of the mass distribution \citep{Abbott:2020gyp}. The tapering function is of the form
\begin{equation}
\label{eq:smoothing}
S(m | m_{\rm min}, \delta_m) = 
\begin{cases}
    0 & \text{for $m< m_{\rm min}$}\\
    f(m - m_{\rm min}, \delta_m) & \text{for $m_{\rm min} \leq m < m_{\rm min}+\delta_m$} \\
    1 & \text{for $m\geq m_{\rm min} + \delta_m$}
\end{cases}
\end{equation}
with
    $f(m, \delta) = \left[1 + \exp \left(\frac{\delta}{m} + \frac{\delta}{m - \delta}\right)\right]^{-1}$.

\vspace{3mm}
For the two mass models listed above, the secondary mass component $m_{2,s}$ is defined as:
\begin{equation}
    p(m_{2,s}|m_{1,s},\mmin,\alpha)=\mathcal{P}(m_{2,s}|\mmin,m_{1,s},\beta).
\end{equation}

\section{Quick generation of posterior samples}
\label{app: b quick posterior samples}

In order to quickly generate posterior samples for the studies in Sec.~\ref{sec:3}-\ref{sec:4}, we use an approach similar to that of \cite{2018ApJ...863L..41F,Farr:2019twy}.

We start by generating the redshift distribution (uniform in comoving volume) and source-frame masses from distribution that we have chosen for the population. For each binary, we calculate the detector-frame chirp mass $\mathcal{M}_d$ and its luminosity distance $d_L$. The optimal SNR of the binary is then taken to be given by 
\begin{equation}
    \rho =  \rho^* \Theta \left(\frac{\mathcal{M}_d}{\mathcal{M^*}_d} \right)^{5/6} \left(\frac{d^*_L}{d_L} \right),
    \label{eq:snr}
\end{equation}
where $\mathcal{M^*}_d$ and $d^*_L$ are a reference chirp mass and luminosity distance at which an optimally oriented binary has optimal SNR $\rho^*$. For our simulation we choose $\mathcal{M^*}_d = 10 M_\odot, d^*_L= 1 {\rm Gpc}$ and  $\rho^*=8$, which are scales compatible with the observing scenario for BBHs during the O3 run \cite{Aasi:2013wya}.  Finally, we assume that the  projection factor $\Theta$ can be drawn from a uniform distribution between $[0,1]$. 

In order to mimic the effect of the noise on the signal recovery, for each binary we draw a detected SNR $\rho_{\rm det}$ from a Gaussian distribution with mean $\rho$ and variance 1. This is the SNR distribution expected in the case of a single detector. Extension to multiple detectors would be straightforward using a $\chi^2$ distribution for the SNR. In our simulation, events are detected if $\rho_{\rm det}$ exceeds a threshold of 12.

Once that we have a list of detected signals (or triggers), we simulate posterior samples. To do so, we first draw the measured chirp mass $\mathcal{M}_{\rm d, det}$ and mass-ratio $q_{\rm d, det}$ from the following likelihoods that approximate the error budgets from full parameter estimation analyses \cite{2016ApJ...825..116F}
\begin{align}
    \mathcal{M}_{\rm d, det} &\propto \mathcal{N}\left(\mathcal{M}_d, 10^{-3} \mathcal{M}_d \frac{10}{\rho_{\rm det}}\right) \\
    q_{\rm d, det} &\propto \mathcal{N}\left(q, 0.25 q \: \frac{10}{\rho_{\rm det}}\right).
\end{align}
Posterior samples on $q$ and $\mathcal{M}_d$ are then generated around the measured values using the above likelihood models. The corresponding values of the detector-frame masses are then given by 
\begin{eqnarray}
    m_{1,d}=\mathcal{M}_d \frac{(1+q)^{1/5}}{q^{3/5}}, \quad m_{2,d}=q m_{1,d}.
\end{eqnarray}
This procedure takes into account the degeneracy between the determination of the two masses.

Finally, in order to account for the degeneracy between luminosity distance and binary inclination angle, we draw a detected projection factor $\Theta_{\rm det}$ from a normal distribution  
\begin{equation}
    \Theta_{\rm det} \propto \mathcal{N}(\Theta, 0.3 \frac{10}{\rho_{\rm det}}),
\end{equation}
and we draw posterior samples on $\Theta$ around this value.

The posterior samples on the luminosity distance are obtained by drawing SNR posterior samples around $\rho_{\det}$ and inverting Eq.~\eqref{eq:snr} using the posterior samples already obtained for $\mathcal{M}_d$ and $\Theta$. This way of generating posterior samples allows to generate samples of the luminosity distance and masses that are consistent with the selection effects accounted for in the analysis.

This results on characteristic uncertainties for the luminosity distance and masses of 40\%-60\% and 20\%-50\% respectively at the 90\% confidence level.

\section{Computational challenges in population analyses \label{app:d}}

Monte Carlo Markov chain (MCMC) algorithms can be employed to sample the posterior of population hyper-parameters. To do so, the MCMC needs to evaluate many times\footnote{The number of iterations depends on the sampling algorithm, the underlying population and the number of events used.} the hierarchical likelihood
\begin{equation}
    p(\{x\}|\Lambda,N_{\rm obs}) \propto \prod_{i}^{N_{\rm obs}} \frac{\int p(x_i|\Lambda,\theta)p_{\rm pop}(\theta|\Lambda)d\theta}{\int p_{\rm det}(\theta,\Lambda)p_{\rm pop}(\theta|\Lambda)d\theta}.
    \label{eq:pos_a}
\end{equation}
This evaluation is computationally demanding for two reasons: \textit{(i)} the GW likelihood is not known analytically and should be computed from posterior samples $p(\theta|x,\Lambda)$ and \textit{(ii)} the likelihood should be evaluated for all events. Thus, calculating Eq.~\eqref{eq:pos_a} in a MCMC looping over the GW events becomes prohibitive for larger number of events.

The evaluation of the denominator that accounts for selection effects is not an issue as its computation can be done once for all the GW events. 

The numerator involves an integral for every GW event considered and for every set of population parameters tried. Using the ``posterior sample recycling'' \cite{Talbot:2019okv} technique the computation of this integral can be efficiently calculated. The integral in the numerator is evaluated as
\begin{equation}
    \int p(x|\Lambda,\theta)p_{\rm pop}(\theta|\Lambda)d\theta \approx \frac{p(x)}{N_s} \sum_{i=1}^{N_s} \frac{p_{\rm pop}(\theta_i|\Lambda)}{\pi(\theta_i|\Lambda)},
    \label{eq:approx}
\end{equation}
where $N_s$ is the number of posterior samples provided from the GW data analysis, $\pi(\theta_i|\Lambda)$ is the original prior applied to generate the posterior samples and $p(x)$ is the evidence computed while sampling the GW posterior (it can be assumed as constant for a fixed waveform and noise model).

Eq.~\eqref{eq:approx} provides an efficient procedure to evaluate the numerator of hierarchical likelihood. 

The ``posterior samples recycling'' is a procedure that can be employed when the GW posterior $p(\theta|x,\Lambda)$ is confined in a smaller volume with respect to the one of $p_{\rm pop} (\theta|\Lambda)$. This is generally true for the current observations of BBHs since their mass and redshift estimates span a smaller range with respect to the population-induced priors for the source-frame masses and redshift distribution. However, as we discuss in the next two sections, this is not the case when the information from EM counterpart or from galaxy surveys is including in the analysis.

\tocless \subsection{\ldots when dealing with EM counterparts}

The redshift information obtained from an EM counterpart can be included by replacing redshift prior with the EM likelihood $p(z_{\rm obs}|z)$ that accounts for the accuracy of the redshift measurement
\begin{equation}
    p_{\rm EM}(z_{\rm obs}|\Lambda)= \int p(z_{\rm obs}|z)\:p_{\rm pop}(z|\Lambda) dz.
\end{equation}

For GW170817, the uncertainty on the redshift from EM counterpart was $\leq 10\%$ (taking into account the uncertainty on the peculiar motion). It was even smaller for the candidate EM counterpart associated with GW190521 \cite{Graham:2020gwr}. The uncertainty obtained on the redshift from GW observations only is much larger, of the order of $25\%-40\%$ which thus forbids the use of posterior sample recycling ``as is'' to compute the integral in Eq.~\ref{eq:approx}.

In order to circumvent this difficulty, a possible approach is to sum over the redshift samples drawn from the EM posterior instead of the GW posterior, as follows:
\begin{equation}
    \int p(x|\Lambda,\theta)p_{\rm pop}(\theta|\Lambda)d\theta \approx  \frac{p(x)}{N_{EM}} n(\Lambda) \sum_{i=1}^{N_{EM}} \frac{p(x|z_i)}{\pi(z_i|\Lambda)},
    \label{eq:approx_EM}
\end{equation}
where $p(x|z_i)$ is an interpolation of the GW likelihood obtained by a kernel density estimate (KDE). The normalization terms $n(\Lambda)$ results from the source-frame mass marginalization by:
\begin{equation}
    n(\Lambda)=\frac{1}{N_s}\sum_{j=1}^{N_s} \frac{p(m^{j}_{1,s}(\Lambda,z_j),m^{j}_{2,s}(\Lambda,z_j)|\Lambda)}{\pi(m^{j}_{1,s}(\Lambda,z_j),m^{j}_{2,s}(\Lambda,z_j)|\Lambda)}.
\end{equation}
and are computed for every $i-th$ value of the population parameters.
Contrarily to the canonical posterior samples recycling this procedure is not parallelizable since the KDE fitting and normalization computation have to be done for every event and every set of population assumptions, thus leading to an unsustainable computational burden.

On top of the above difficulty, the selection bias due to the EM detection also have to be modelled as they can play an important r\^ole especially for BNS \cite{Chen:2020dyt,Mastrogiovanni:2020ppa} but also for BBH \cite{2021arXiv210316069P} if the EM counterpart of GW190521 is confirmed.

\tocless \subsection{\ldots when using galaxy catalogs}

The inclusion of galaxy catalogs to the analysis can be done by replacing the redshift prior by the distribution of galaxies obtained from a survey.

However, this approach relies on the completeness of the galaxy catalog (no matter the source) and contains with $100\%$ probability the hosting galaxy of the GW event. If this is not the case, then a selection bias could be introduced and one needs to correct it for the galaxy catalog incompleteness, see \cite{Gray:2019ksv} for more details. The completeness correction is itself a non-trivial function (through the galaxy luminosity distributions) of the cosmological parameters and its computation adds a significant burden to the load of the analysis, since galaxy catalogs are usually composed by billions of data points.
Thus, a complete population inference using galaxy catalogs will require developments in terms of data analysis and computing techniques.

\section{Hierarchical posterior with EM counterparts}
\label{app:scream}

In this Appendix we calculate the hierarchical posterior when, for all events, an EM counterpart provides a redshift measurement. We begin from the hierarchical posterior in Eq.~\eqref{eq:EMcount}

\begin{align*}
p(\Lambda_c,\Lambda_m|x,z_{\rm obs}) & \propto p(\Lambda_c,\Lambda_m) p(x,z_{\rm obs}|\Lambda_c,\Lambda_m)\\
& = p(\Lambda_c,\Lambda_m) \frac{\int p(x|\Lambda,\theta) p (z_{\rm obs}|z,\bar{\theta}) p_{\rm pop}(\theta|\Lambda)d\theta}{\int p_{\rm det,GW}(\theta,\Lambda) p_{\rm det,EM}(\theta,\Lambda)p_{\rm pop}(\theta|\Lambda)d\theta}.
\end{align*}

We  assume that the redshift measurement is accurate and does not depend upon the binary parameters, i.e.~$p(z_{\rm obs}|z,\bar{\theta})\approx \delta(z_{\rm obs}-z)$. 
This may be an over-simplication, especially for BNSs for which the detection of the EM counterpart can be strongly related to inclination of the orbital plane with the line-of-sight, see 
\cite{Mastrogiovanni:2020ppa,2020A&A...639A..15D, Chen:2020zoq} for more details.

We also assume that the measurements of the luminosity distance $d_L$ and detector frame masses are mutually independent, i.e. $p(x|d_L,m_{1,d},m_{2,d}) \propto p(x|d_L) p(x|m_{1,d},m_{2,d})$.
This is a reasonable assumption given that the estimate of the masses comes primarily from the GW phase while the estimate of the luminosity distance comes from the amplitude.
These likelihood terms can be evaluated as
\begin{align}
    p(x|d_L) &\propto \frac{p(d_L|x)}{\pi(d_L)} \\
    p(x|m_{1,d},m_{2,d}) &\propto \frac{p(m_{1,d},m_{2,d}|x)}{\pi(m_{1,d},m_{2,d})},
\end{align}
where $\pi(\cdot)$ are the priors used to generate the posterior samples. With these assumptions, the hierarchical posterior becomes
\begin{eqnarray}
     p(\Lambda_c,\Lambda_m|x,z_{\rm obs}) \propto  \frac{\pi(\Lambda)}{p(\mathscr{D}|\Lambda)}  p(x|d_L(\Lambda_c,z_{\rm obs})) p(z^{i}_{\rm obs}|\Lambda_c) I(x;\Lambda_m,z_{\rm obs}),
     \label{eq:EM_simp}
\end{eqnarray}
where we have defined the integral function $I$ as 
\begin{eqnarray}
    I(x;\Lambda_m,z_{\rm obs}) = \int p(x|m_{1,d} (m_{1,s},z_{\rm obs}),m_{2,d}(m_{2,s},z_{\rm obs})) p_{\rm pop}(m_{1,s},m_{2,s}|\Lambda_m) dm_{1,s} dm_{2,s}.
    \label{eq:Idef}
\end{eqnarray}

Eq.~\eqref{eq:EM_simp} evidences a important result: under the two assumptions formulated above, the estimation of the population parameters $\Lambda_m$ does not impact the estimation of the cosmological parameters $\Lambda_c$ through the parameters measured from the GW. The constraints on the cosmological and population parameters come from distinct terms in the above equation,
respectively, on $p(x|d_L(\Lambda_c,z_{\rm obs}))$ and $I(x;\Lambda_m,z_{\rm obs})$.

This remains true when the population prior on the source-frame masses depends on the redshift. The only correlation between $\Lambda_c$ and $\Lambda_m$ lies in the selection effect term $p(\mathscr{D}|\{\Lambda_c, \Lambda_m\})$ and in the joint prior $\pi(\{\Lambda_c, \Lambda_m\})$.
Eq.~\eqref{eq:EM_simp} can be easily extended to a population of GW sources, 
\begin{equation}
    p(\Lambda_c,\Lambda_m|\{x\},z_{\rm obs}) \propto \frac{\pi(\Lambda)}{[p(\mathscr{D}|\Lambda)]^{N_{\rm obs}}}  \prod_i^{N_{\rm obs}}  p(z^{i}_{\rm obs}|\Lambda_c) p(x^i|d_L(\Lambda_c,z^{i}_{\rm obs})) \prod_i^{N_{\rm obs}}  I(x^{i};\Lambda_m,z^{i}_{\rm obs}).
\end{equation}

\twocolumngrid

\bibliographystyle{utphys}
\bibliography{refs}

\end{document}